\documentclass[acmsmall]{acmart}
\pdfoutput=1
\usepackage{booktabs} 
\usepackage{subcaption}
\usepackage[ruled, linesnumbered]{algorithm2e}
\usepackage{amsmath}
\usepackage{multicol}
\usepackage{graphicx}
\usepackage[noend]{algpseudocode}
\usepackage{textcomp}
\usepackage{flexisym}
\usepackage{enumitem}
\usepackage{stackengine}
\usepackage{booktabs}
\usepackage{multirow}
\usepackage{wrapfig}

\newcommand{\skm}[1]{\textcolor{black}{#1}} 

\newcommand{\revfinal}[1]{\textcolor{black}{#1}} 
\newcommand{\camready}[1]{\textcolor{black}{#1}} 

\begin{document}
 \setcopyright{acmcopyright}
 \acmJournal{TECS}
 \acmYear{2019} \acmVolume{1} \acmNumber{1} \acmArticle{1} \acmMonth{1} \acmDOI{https://doi.org/10.1145/3126530}
\title{\revfinal{Analytical Performance Models for NoCs with Multiple Priority Traffic Classes}}

\author{Sumit K. Mandal}
\orcid{0000-0002-9294-1603}
\affiliation{%
\institution{Arizona State University}
\department{School of Electrical, Computer and Energy Engineering}
\city{Tempe}
\state{AZ}
\postcode{85287}
\country{USA}
}
\author{Raid Ayoub}
\affiliation{%
\institution{Intel Corporation}
\city{Hillsboro}
\state{OR}
\postcode{97124}
\country{USA}
}
\author{Michael Kishinevsky}
\affiliation{%
\institution{Intel Corporation}
\city{Hillsboro}
\state{OR}
\postcode{97124}
\country{USA}
}
\author{Umit Y. Ogras}
\affiliation{%
\institution{Arizona State University}
\department{School of Electrical, Computer and Energy Engineering}
\city{Tempe}
\state{AZ}
\postcode{85287}
\country{USA}
}
\email{{skmandal,umit}@asu.edu,  {raid.ayoub,michael.kishinevsky}@intel.com}

\begin{abstract}
Networks-on-chip (NoCs) have become the standard for interconnect solutions in industrial designs ranging from client CPUs to many-core chip-multiprocessors. 
Since NoCs play a vital role in system performance and power consumption, 
pre-silicon evaluation environments include cycle-accurate NoC simulators.
Long simulations increase the execution time of evaluation frameworks, 
which are already notoriously slow, and prohibit design-space exploration. 
Existing analytical NoC models, which assume fair arbitration, cannot replace these simulations since industrial NoCs typically employ priority schedulers and multiple priority classes.
To address this limitation, 
we propose a systematic approach to construct priority-aware analytical performance models using micro-architecture specifications and input traffic. 
Our approach decomposes the given NoC into individual queues with modified service time to enable accurate and scalable latency computations. Specifically, we introduce novel transformations along with an algorithm that iteratively applies these transformations to decompose the queuing system.
Experimental evaluations using real architectures and applications show high accuracy of 97\% and up to 2.5$\times$ speedup in full-system simulation.	
\end{abstract}

\ccsdesc[500]{Networks~Network performance modeling}
\ccsdesc[500]{Computer systems organization~System on a chip}
%
\thanks{This article will appear as part of the ESWEEK-TECS special issue and will be presented in the International Conference on Compilers, Architecture, and Synthesis for Embedded Systems (CASES) 2019


Author's addresses: S. K. Mandal {and} U. Y. Ogras, School of
Electrical,
Computer and Energy Engineering, Arizona State University, Tempe, AZ,
85287; emails: \{skmandal, umit\}@asu.edu;
Raid Ayoub {and} Michael Kishinevsky, Intel Corporation, 2111 NE 25th Ave.,
Hillsboro, OR 97124; emails: \{raid.ayoub, michael.kishinevsky\}@intel.com
}
\maketitle

\section{Introduction} \label{sec:intro}

Modern design methodologies in industries involve thorough power, performance, and area evaluations before the architectural decisions are frozen. 
These pre-silicon evaluations are vital for detecting functional bugs and power-performance violations, since post-silicon fixes are costly, if feasible at all. 
Therefore, a significant amount of resources are dedicated to pre-silicon evaluation 
using virtual platforms \cite{leupers2011virtual} or full-system simulators \cite{Binkert2011Gem5}.
NoC simulations play a critical role in these evaluations, 
as NoCs have become the standard interconnect solution in many core chip-multiprocessors (CMPs) \cite{jeffers2016intel,keltcher2003amd,kahle2005introduction}, client CPUs~\cite{rotem2015intel}, and mobile systems-on-chip \cite{singh2014evolution}. 
\camready{Moreover, there is a growing interest to use NoCs in hardware implementations of deep neural networks~\cite{choi2017chip}.}

\begin{figure}[t]
	\centering
	\vspace{-4mm}
	\includegraphics[width=0.6\columnwidth]{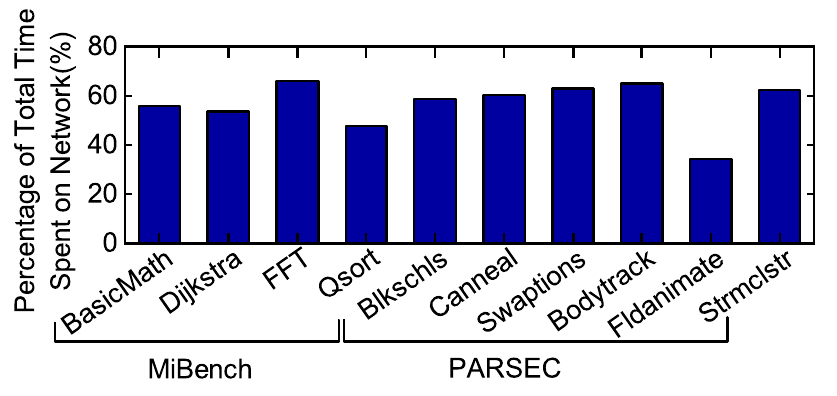}
	\vspace{-4mm}
	\caption{Experiments of different applications show that 40\%-70\% of the total simulation time is spent on the network.} 
	\label{fig:gem5_profile}
	\vspace{-4mm}
\end{figure}

Since the on-chip interconnect is a critical component of multicore architectures, pre-silicon evaluation platforms contain cycle-accurate NoC simulators~\cite{agarwal2009garnet, jiang2013detailed}. 
NoC simulations take up a significant portion of the total simulation time, 
which is already limiting the scope of pre-silicon evaluation (e.g., 
simulating even a few seconds of applications can take days).
For example, Figure \ref{fig:gem5_profile} shows that 40\%-70\% of total simulation time is spent on the network itself when performing full-system simulation using gem5~\cite{Binkert2011Gem5}.
Hence, accelerating NoC simulations without sacrificing accuracy can significantly improve both the quality and scope of pre-silicon evaluations.
%
%
\begin{figure}[b]
	\centering
	\includegraphics[width=0.6\columnwidth]{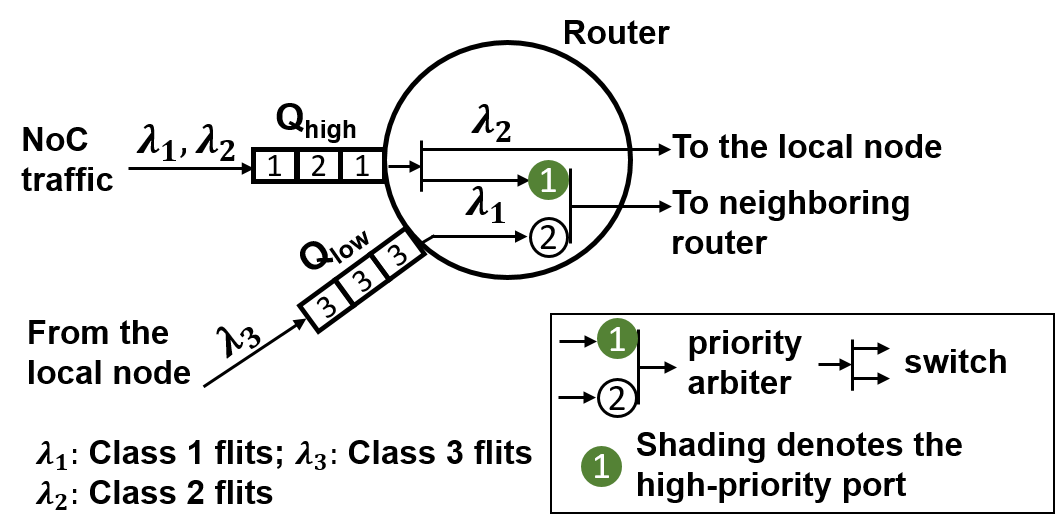}
	\caption{The high priority queue (Q$_{\mathrm{high}}$) stores two different traffic classes which are already in the NoC, while the low priority queue (Q$_{\mathrm{low}}$) stores the newly injected flits from the local node. As flits from class--2 are routed to the local node, low-priority flits compete with only class--1 flits in Q$_{\mathrm{high}}$.} 
	\label{fig:motivation}
	\vspace{-3.5mm}
\end{figure}

Several performance analysis approaches have been proposed 
to enable faster NoC design space exploration~\cite{ogras2010analytical,wu2010analytical,qian2015support}.  
Prior techniques have assumed a round-robin arbitration policy in the routers since the majority of router architectures proposed to date have used round-robin for fairness.
In doing so, \textit{they miss two critical aspects} of the industrial priority-based NoCs~\cite{jeffers2016intel,rico2017arm,kahle2005introduction}.
First, routers give priority to the flits in the network to achieve predictable latency within the interconnect. 
\camready{For example, let us assume, class-1 flits to the neighboring router and class-2 flits to the local node in Figure~\ref{fig:motivation} 
are already in the NoC, while flits from class-3 to the neighboring router must wait in the input buffer to be admitted.}
Consequently, flits in the NoC (class-1 and class-2) experience deterministic service time at the expense of increased waiting time for new flits. 
Second, flits from different priority classes can be stored in the same queue. 
For instance, new read/write requests from the core to tag directories use the same physical and virtual channels as the requests forwarded from the directories to the memory controllers. 
Moreover, only a fraction of the flits in either the high or low 
priority queue can compete with the flits in the other queue. 
For example, suppose the class-2 flits in Figure~\ref{fig:motivation} 
are delivered to the local node. 
Then, class-3 flits must compete with only class-1 flits in the high-priority queue. 
Analytical models that ignore this traffic split significantly overestimate the latency,
as shown in Section~\ref{sec:sp_hp}. 
In contrast, analytical models that ignore priority would significantly 
underestimate the latency. 
Thus, prior approaches that do not model priority~\cite{ogras2010analytical,wu2010analytical,qian2015support}
and simple performance models for the priority queues~\cite{kiasari2013analytical,jin2009modelling} 
are \textit{inapplicable to priority-based industrial NoCs}.

This paper presents a novel NoC performance analysis technique that \textit{considers traffic classes with different priorities}. 
This problem is theoretically challenging due to the non-trivial interactions between classes and shared resources. 
For example, queues can be shared by flits with different priorities, as shown in Figure~\ref{fig:motivation}. 
Similarly, routers may \textit{merge} different classes coming through separate ports, 
or act as \textit{switches} that can disjoin flits coming from different physical channels.
To address these challenges, we propose a two-step approach that consists of an analysis technique followed by an iterative algorithm.  
The first step establishes that priority-based NoCs can be decomposed into separate queues using traffic splits of two types. 
Since existing performance analysis techniques cannot model these structures with traffic splits, we develop analytical models for these canonical queuing structures. 
The second step involves a novel iterative algorithm that composes an end-to-end latency model for the queuing network of a given NoC topology and input traffic pattern. 
The proposed approach is evaluated thoroughly using both 2D mesh and ring architectures used in industrial NoCs.
It achieves 97\% accuracy with respect to cycle-accurate simulations for realistic architectures and applications.

\noindent \textit{The major contributions of this paper are as follows:} 
\vspace{-0.5mm}
\begin{enumerate}  [leftmargin=9mm]
	\item A technique to obtain analytical performance models of priority-based NoCs with multiple traffic classes,
	\item An algorithm to synthesize end-to-end latency for each traffic class using the analytical models, 
	\item Extensive evaluations using an industrial cycle-accurate simulator, real applications and synthetic traffic.
\end{enumerate}
The rest of the paper is organized as follows. 
Section~\ref{sec:related_work} reviews the related work.
Section~\ref{sec:overview} provides a brief overview and background of the proposed work.
Section~\ref{sec:methodology} describes the required transformations for two canonical structures and their analytical models. 
Section~\ref{sec:model_gen} describes how these two transformations are used to analyze a queuing network. 
Section~\ref{sec:exp_eval} presents experimental evaluations, and 
Section~\ref{sec:concl_future} concludes the paper summarizing the key contributions.

%

%
 

\vspace{-0.25mm}
\section{Related Work} \label{sec:related_work}
\vspace{-0.25mm}

Performance analysis techniques are useful for exploring design space~\cite{pande2005performance} and speeding up simulations ~\cite{ogras2010analytical,kiasari2013analytical,wu2010analytical}. 
Indeed, there is continuous interest in applying novel techniques such as machine learning~\cite{qian2015support} 
and network calculus~\cite{qian2009analysis} to NoC performance analysis.
However, these studies do not consider multiple traffic classes with different priorities.
Since state-of-the-art industrial NoC designs~\cite{doweck2017inside, jeffers2016intel} use 
\textit{priority-based arbitration} with multi-class traffic, it is important to develop performance analysis for this type of architectures. 

Kashif et al. have recently presented priority-aware router architectures~\cite{kashif2014bounding}.
However, this work presents analytical models only for worst-case latency. 
In practice, analyzing the average latency is important 
since using worst-case latency estimates in full-system would lead to inaccurate conclusions. 
A recent technique proposed an analytical latency model for priority-based NoC~\cite{kiasari2013analytical}.
This technique, however, assumes that each queue in the network contains a single class of flits.

Several techniques present performance analysis of priority-based queuing networks outside the NoC domain~\cite{bertsekas1992data,bolch2006queueing,ikeharaapproximate}.
Nevertheless, these techniques do not consider multiple traffic classes in the same queue. 
The work presented in~\cite{awan2005analysis} considers multiple traffic classes, 
but it assumes that high priority packets preempt the lower priority packets. 
However, this is not a valid assumption in the NoC context.
A technique that can handle two traffic classes, Empty Buffer Approximation (EBA), 
has been proposed in \cite{berger2000workload} for a priority-based queuing system. 
This approach was later extended to multi-class systems~\cite{jin2009modelling}. 
However, EBA ignores the residual time caused by low priority flits on high priority traffic. 
Hence, it is impractical to use EBA for priority-aware industrial NoCs.

The aforementioned prior studies assume a continuous-time queuing network model, 
while the events in synchronous NoCs take place in discrete clock cycles. 
A discrete-time priority-based queuing system is analyzed in~\cite{walraevens2004discrete}.
This technique forms a Markov chain for a given queuing system, 
then analyzes this model in z-domain through probability generating functions (PGF).
PGFs deal with joint probability distributions where the number of random variables is equal to the number of traffic classes in the queuing system. 
This approach is not scalable for systems with large number of traffic classes because the corresponding analysis becomes intractable.
For example, an industrial 8$\times$8 NoC would have 64 sources and 64 destinations which will result in 4096 (64$\times$64) variables with PGF.
Furthermore, our approach outperforms this technique, as demonstrated in Section~\ref{sec:mesh_result}.

In contrast to prior approaches, we propose a scalable and accurate closed form solution for a priority-based queuing network with multi-class traffic.
The proposed technique constructs end-to-end latency models using two canonical structures identified for priority-based NoCs.
Unlike prior approaches, our technique scales to any number of traffic classes.
\camready{To the best of our knowledge, this is the first analytical model for priority-based NoCs that considers both (1) shared queues among multiple priority classes and (2) traffic arbitration dependencies across the queues.}
\section{Overview and Background} \label{sec:overview}






\begin{figure}[b]
	\centering
	\vspace{-3mm}
	\includegraphics[width=0.6\columnwidth]{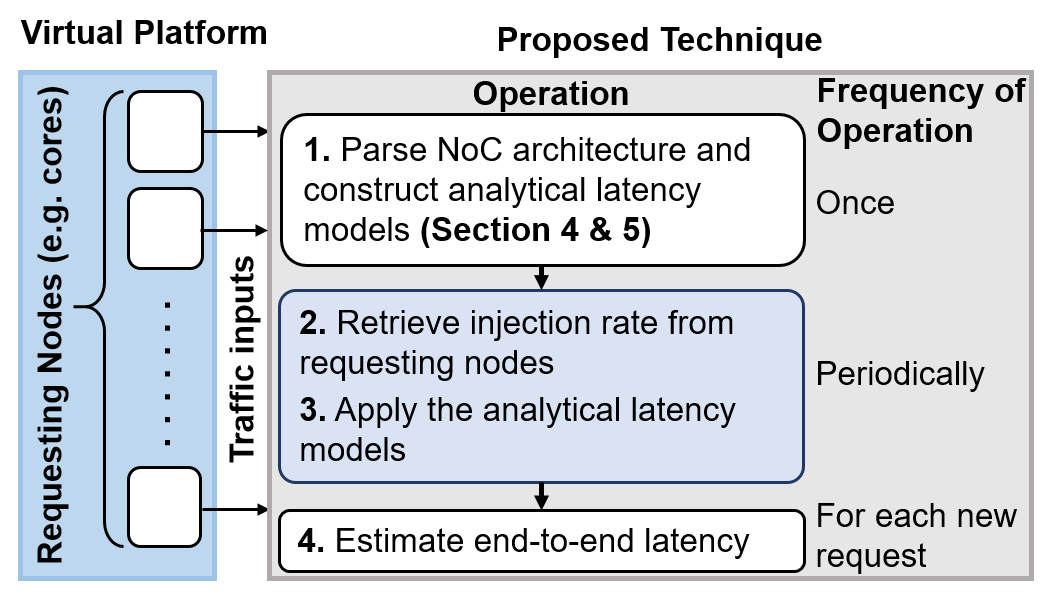}
	\vspace{-4mm}
	\caption{Overview of the proposed methodology.} 
	\vspace{-1mm}
	\label{fig:overview}
\end{figure}

\subsection{Proposed Performance Analysis Flow} \label{sec:use_model}
The primary target of the proposed model is to accelerate virtual platforms~\cite{bartolini2010virtual} and full-system simulations~\cite{Binkert2011Gem5,patel2011marss,magnusson2002simics}
by replacing time-consuming NoC simulations with accurate lightweight analytical models.
\camready{At the beginning of the simulation, the proposed technique parses the priority-based NoC topology to construct the analytical models, as shown in Figure~\ref{fig:overview}.}
The host, such as a virtual platform, maintains a record of traffic load and the destination address for each node. 
It also periodically (each 10K-100K cycles) sends the traffic injections of requesting nodes, such as cores, 
to the proposed technique. 
Then, the proposed technique applies the analytical models (steps 2 and 3 in Figure~\ref{fig:overview}) to compute the end-to-end latency.
Whenever there is a new request from an end node, the host system estimates the latency using the proposed model as a function of the source-destination pair. 
That is, our model replaces the cycle-by-cycle simulation of flits in NoCs.


We note that the requesting nodes, such as the cores, 
have a parameterized number of maximum outstanding request credits. 
Hence, the requesters are automatically throttled by these credits/node, leading to a bounded number of flits in the NoCs. 
Since the traffic injection rates provided by the host already account for this throttling, 
we do not explicitly model blocking at the interfaces.





\subsection{Basic Priority-Based Queuing Models}

We assume a discrete time system in which micro-architectural events, 
such as writing to a buffer, arbitration and switch traversal happen in the integral number of clock cycles. 
Therefore, we develop queuing models based on arrival process that follows geometric distribution, 
\textit{in contrast to} continuous time models that are based on Poisson (M for Markovian) arrival assumption. 
More specifically, we adopt the Geo/G/1 model, in which the inter-arrival time of the incoming flits to the queue follows geometric distribution (denoted by Geo), service time of the queue follows a general discrete-time distribution (denoted by G), and the queue has one server (the `1' in the Geo/G/1 notation).
The proposed technique estimates the end-to-end latency for realistic applications accurately, as we demonstrate in Section~\ref{sec:real_apps}. 
However, the accuracy is expected to drop if the NoC operates close to its maximum load since the Geometric (similar to Poisson) packet inter-arrival time assumption becomes invalid~\cite{ogras2010analytical}.

\begin{figure}[t]
 	\vspace{-2mm}
	\centering
	\resizebox{0.6\textwidth}{!}{\includegraphics[width=0.5\textwidth]{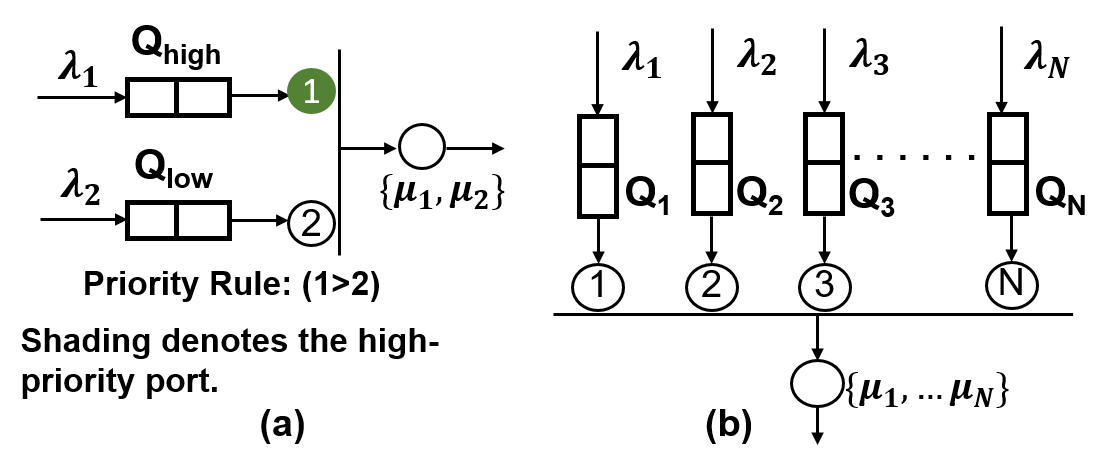}}
	\caption{(a) A system with two queues. Flits in Q$_{high}$ have higher priority than flits in Q$_{low}$. 
	(b) A system with N queues, 
	where Q$_{i}$ has higher priority than Q$_{j}$ for $i<j$}
	\vspace{-6mm}
	\label{fig_basic_priority}	
\end{figure}

Performance analysis techniques in the literature~\cite{bertsekas1992data,jin2009modelling,kiasari2013analytical} discuss basic priority-based networks in which each priority class has a dedicated queue, as illustrated in Figure~\ref{fig_basic_priority}(a). 
In this architecture, the flits in Q$_{high}$ have higher priority than the flits in Q$_{low}$. 
That is, flits in Q$_{low}$ will be served \textit{only when} Q$_{high}$ is empty and the server is ready to serve new flits. 
Another example with $N$ priority classes is shown in Figure~\ref{fig_basic_priority}(b).
The flits in Q$_i$ have higher priority than flits in Q$_{j}$ if $i<j$. 
The average waiting time for each priority class $W_i$ for $1 \leq i \leq N$ is known for continuous time M/G/1 queues~\cite{bertsekas1992data}.
In the M/G/1 queuing system, flits arrive in the queue following Poisson distribution (M) and the service time of the queue follows general distribution (G).
In this work, we first derive waiting time expressions for discrete time Geo/G/1 queues. 
Then, we employ these models to derive end-to-end NoC latency models.


The average waiting time of flits in a queue can be divided into two parts: 
(1) waiting time due to the flits already buffered in the queue, and
(2) waiting time due to the flits which are in the middle of their service, i.e., the residual time. 
The following lemma expresses the waiting time as a function of input traffic and NoC parameters.

\noindent\textbf{Lemma~1:} Consider a queuing network with $N$ priority classes as shown in Figure~\ref{fig_basic_priority}(b). 
Suppose that we are given the injection rates $\lambda_i$, service rates $\mu_i$, residual time $R_i$, and server utilizations $\rho_i$ for $1 \leq i \leq N$, where $N \geq 2$ (see Table~\ref{tab_symbols_used}). 
Then, the waiting time of class-$i$ flits $W_i$ is given as:
\begin{equation} \label{eq:W_basic_prior}
W_i = 
\begin{cases}
\frac{\sum_{k=1}^{N}R_k} {1-\rho_1}, & \text{for $i = 1$ } \\
\frac{\sum_{k=1}^{N}R_k + \sum_{k=1}^{i-1} \Big(\rho_k + \rho_k W_k \Big)}{1 - \sum_{k=1}^{i} \rho_k}, & \text{for $i > 1$}
\end{cases}
\end{equation}%

\noindent\textbf{Proof:} Equation~\ref{eq:W_basic_prior} is derived in Appendix A to avoid distorting the flow of the paper.


\vspace{2mm}
\noindent\textbf{{Shortcomings of the Basic Priority-Based Queuing Models:}} 
Although Equation~\ref{eq:W_basic_prior} extends the known results from continuous time to discrete time queuing systems, 
it cannot handle a network of queues in which each queue can store more than one priority class. 
For example, it does not handle the scenario in which both class-1 and class-2 flits can use Q$_{high}$ in Figure~\ref{fig:motivation}. 
To this end, we present our novel technique that handles multiple priority classes in one queue.

\begin{table}[t]
\caption {Summary of the notations used in this paper}
	\vspace{-3mm}
	\label{tab_symbols_used}
	\begin{tabular}{|l|l|}
\hline
$\lambda_i$       & Injection rate of class-i flits                                                                                                    \\ \hline
$T_i$            & Service time of class-i flits                                                                                                      \\ \hline
$\mu_i$          & Service rate of class-i flits (=1/$T_i$)                                                                                           \\ \hline
$R_i$             & Residual time of class-i flits                                                                                                     \\ \hline
$\rho_i$         & Server utilization of class-i flits (=$\lambda_i T_i$)                                                                             \\ \hline
$C_A$, $C_{A_i}$ & \begin{tabular}[c]{@{}l@{}}Coefficient of variation of inter-arrival time\\ for total flow and class-i respectively\end{tabular}   \\ \hline
$C_B$            & Coefficient of variation of service time                                                                                           \\ \hline
$C_D$, $C_{D_i}$ & \begin{tabular}[c]{@{}l@{}}Coefficient of variation of inter-departure time\\ for total flow and class-i respectively\end{tabular} \\ \hline
$W_i$            & Waiting time of class-i flits                                                                                                      \\ \hline
\end{tabular}
\vspace{-3mm}
\end{table}


\section{Canonical Analytical Models} \label{sec:methodology}

This section describes two canonical queuing structures observed in priority-based NoCs. 
We first describe these structures and explain why prior analysis techniques fail to analyze them.  
Then, we present two novel transformations and accurate analysis techniques. 

\subsection{\hspace{-3mm} Transformation~1: Split at \textit{High} Priority Queue} \label{sec:sp_hp}
\begin{figure*}[b]
	\centering
	\resizebox{1\textwidth}{!}{\includegraphics[width=1\textwidth]{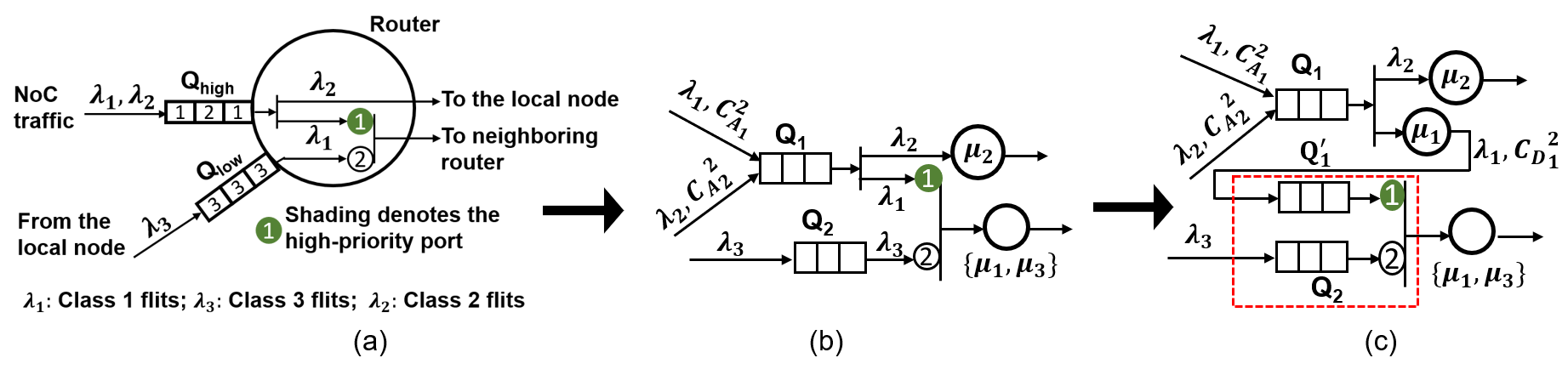}}
 	\vspace{-9mm}
	\caption{Split at high priority: Structural Transformation.}
	\label{fig_strctr_transform}	
\end{figure*}

\begin{figure}[h]
	\centering
	\resizebox{0.6\columnwidth}{!}{\includegraphics[width=0.5\textwidth]{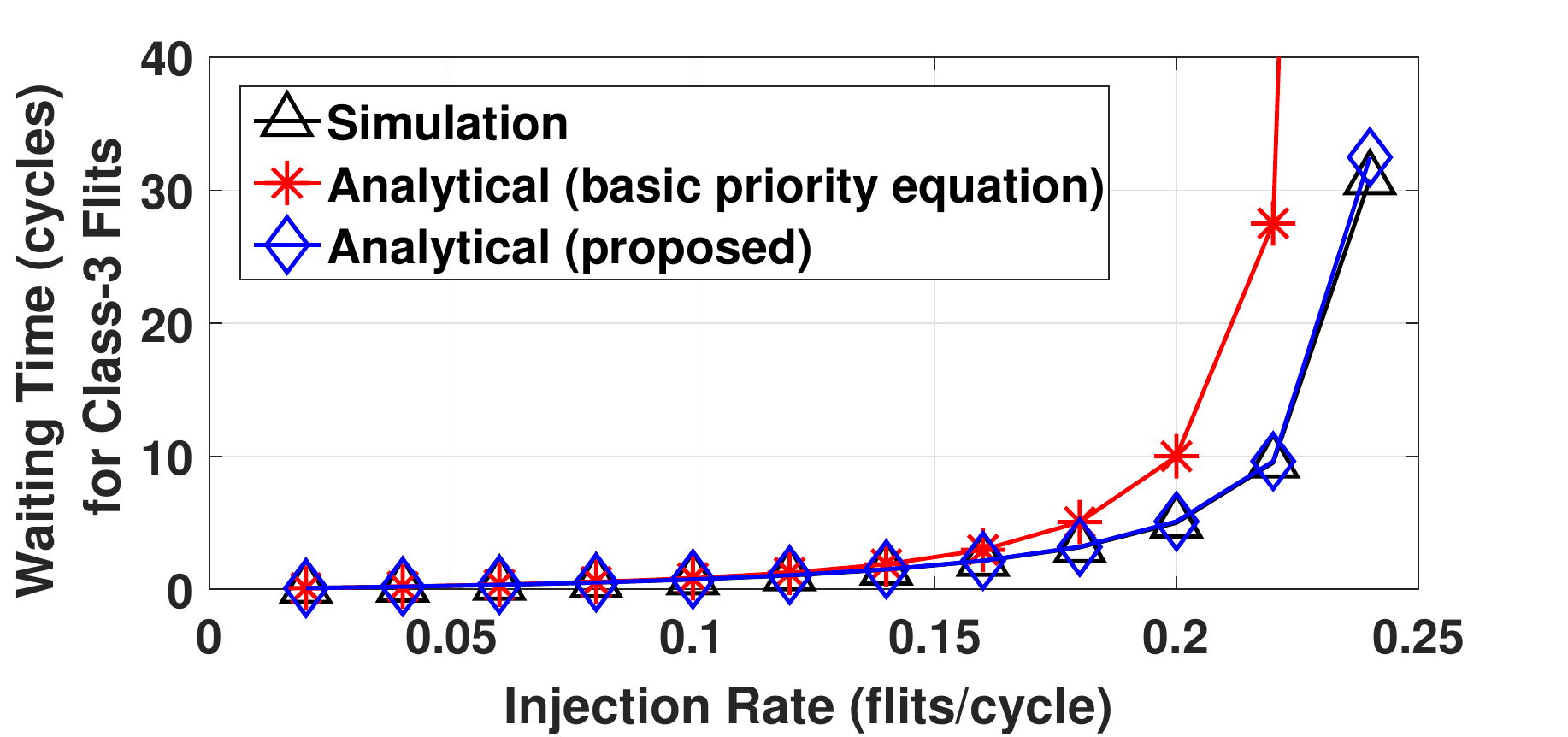}}
	\vspace{-3mm}
    \caption{\camready{Comparison of simulation with the basic priority-based queuing model and proposed analytical model.}}
	\vspace{-3mm}
	\label{fig:sp_hp_basic_comp}	
\end{figure}

\noindent \textbf{Conceptual Illustration:}
Consider the structure shown in Figure~\ref{fig_strctr_transform}(a). 
As illustrated in Section~\ref{sec:intro}, flits from traffic class-1 and 2 are already in the network, while flits from traffic class-3 are waiting in Q$_{low}$ to be admitted.
Since routers give priority to the flits in the network in industrial NoCs, 
class-1 flits have higher priority than those in Q$_{low}$.
To facilitate the description of the proposed models, 
we represent this system by the structure shown in Figure~\ref{fig_strctr_transform}(b).
In this figure, $\mu_i$ represents the service rate of class-$i$ for $i = 1,2,3$.
If we use Equation~\ref{eq:W_basic_prior} to obtain an analytical model for the waiting time of traffic class-3, the resulting waiting time will be highly pessimistic, as shown in Figure~\ref{fig:sp_hp_basic_comp}. 
The basic priority-based queuing model overestimates the latency, since it assumes each class in the network occupy separate queues.
Hence, all flits in Q$_1$ have higher priority than those in Q$_2$.

\noindent \textbf{Proposed Transformation:}
The basic priority equations cannot be applied to this system
since flit distribution of class-1 as seen by class-3 flits will change depending on the presence of class-2 traffic. To address this challenge, we propose a novel structural transformation, Figure~\ref{fig_strctr_transform}(b) to Figure~\ref{fig_strctr_transform}(c). 
Comparison of the structures before and after the transformation reveals: 
\begin{itemize} [leftmargin=*]
    \item The top portion (Q$_1$ with its server) is identical to the original structure, 
    since $\mu_1$ and $\mu_2$ remain the same due to higher priority of class-1 over class-3.
    
    \item The bottom portion (Q$_1\textprime$ and Q$_2$) forms a basic priority queue structure, as highlighted by the red dotted box. 
\end{itemize}
The basic priority queue structure is useful since we have already derived its waiting time model in Equation~\ref{eq:W_basic_prior}. 
However, the arrival process at Q$_1\textprime$ must be derived to apply this equation 
and ensure the equivalence of the structures before and after the transformation.

\begin{figure}[h]
	\vspace{-3mm}
	\centering
	\resizebox{0.6\textwidth}{!}{\includegraphics[width=0.5\textwidth]{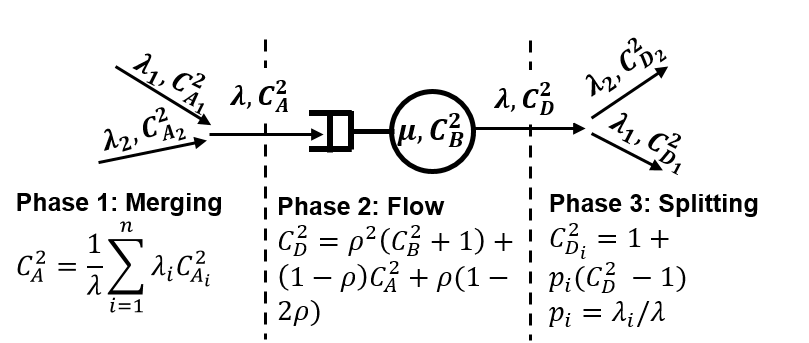}}
	\vspace{-2mm}
	\caption{Decomposition technique: In phase 1, different traffic flows merge into a single flow with an inter-arrival time $C_A$; in phase 2, flits flow into the queue and leave the queue with an inter-departure time $C_D$; in phase 3, flits split into different flows with individual inter-departure time.}
	\label{fig_decomp_tech}	
	\vspace{-4mm}
	
\end{figure}

We derive \camready{the second order moment of inter-departure time of class-1}
using the decomposition technique presented in~\cite{bolch2006queueing}. 
These inter-departure distributions are functions of inter-arrival distributions of all traffic classes flowing in the same queue and service rate of the classes, as illustrated in Figure~\ref{fig_decomp_tech}. 
This technique first calculates the effective coefficient of variation at the input ($C_A^2$)
as the weighted sum of the coefficient of variation of individual classes ($C_{A_i}^2$ in Figure~\ref{fig_decomp_tech}-Phase 1). 
Then, it finds the effective coefficient of variation for the inter-departure time ($C_D^2$) 
using $C_A^2$ and the coefficient of variation for the service time ($C_B^2$). 
In the final phase, the coefficient of variation for inter-departure time of individual classes is found, as illustrated in Figure~\ref{fig_decomp_tech}(Phase 3).  
By calculating the first two moments of the inter-arrival statistics of Q$_1\textprime$ as $\lambda_1$ and $C_{D_1}^2$, 
we ensure that the transformed structure in Figure~\ref{fig_strctr_transform}(c) 
approximates the original system. 
This decomposition enables us to find the residual time for class-1 $R_1^{Q_1\textprime}$ as:
\begin{equation} \label{eq:R_1_Q2}
R_1^{Q_1\textprime} = \frac{1}{2} \frac{\rho_{1}}{\mu_1}  \Bigg(\frac {C_{D_1}^2 + C_{B}^2}{2}\Bigg) - \frac{\rho_1 \mu_1}{2}
\end{equation}

\noindent \textbf{Proposed Analytical Model:}
%
%
The bottom part of the transformed system in Figure~\ref{fig_strctr_transform}(c) 
is the basic priority queue (marked with the dotted red box). 
Therefore, the higher priority part of Equation~\ref{eq:W_basic_prior} can be used to express the waiting time of class-1 flits as:
\begin{equation} \label{eq:W_1_Q2}
W_1^{Q_1\textprime} = \frac {R_1^{Q_1\textprime} + R_{3}} {1 - \rho_1}
\end{equation}
where the residual time of class-1 flits $R_1^{Q_1\textprime}$ is found using Equation~\ref{eq:R_1_Q2}.
Subsequently, this result is substituted in the lower priority portion of Equation~\ref{eq:W_basic_prior} 
to find the waiting time for class-3 flits:
\begin{equation} \label{eq:W_2_Q2}
W_{3} = \frac {R_1^{Q_1\textprime} + R_{3} + \rho_1 + \rho_1 W_1^{Q_1\textprime}} {1 - \rho_1 - \rho_3}
\end{equation}
We also note the waiting time of class-2 flits, $W_2$, is not affected by this transformation. 
Hence, we can express it as $W_2 = \frac{R_2 + R_1}{1-\rho_2}$, using Equation~\ref{eq:W_basic_prior} for the degenerate case of $N=1$.


\begin{figure*}[b]
	\vspace{-5mm}
	\centering
	\resizebox{1\textwidth}{!}{\includegraphics[width=1\textwidth]{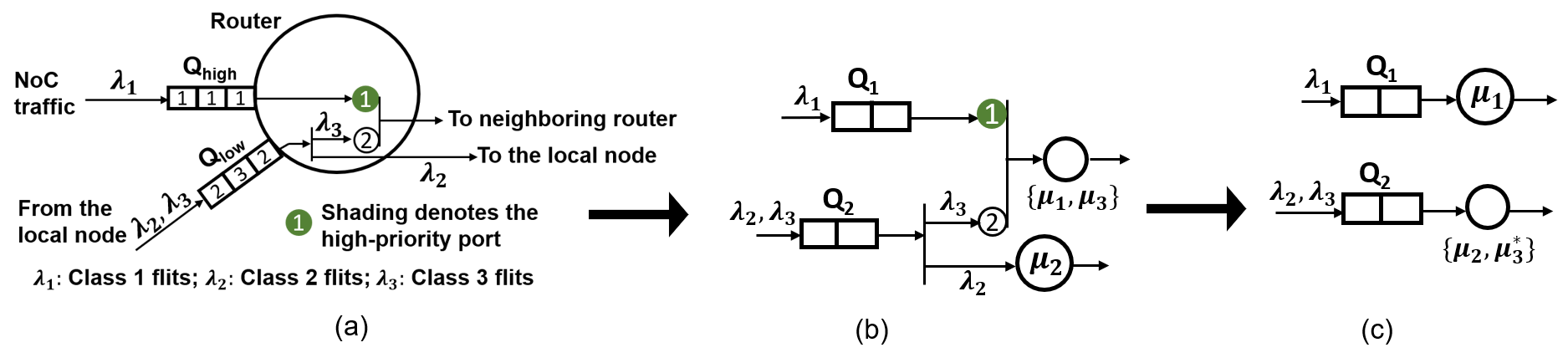}}
	\vspace{-9mm}
	\caption{Split at low priority: Service Rate Transformation. $\mu^*$ denotes transformed service rate. The waiting time of class-1 flits depends on the residual time of the class-3 flits, as shown in Equation~\ref{eq:hp_waiting}.}
	\label{fig:servicerate_transform}
\end{figure*}
\begin{figure}[h]
	\centering
	\resizebox{0.6\textwidth}{!}{\includegraphics[width=0.5\textwidth]{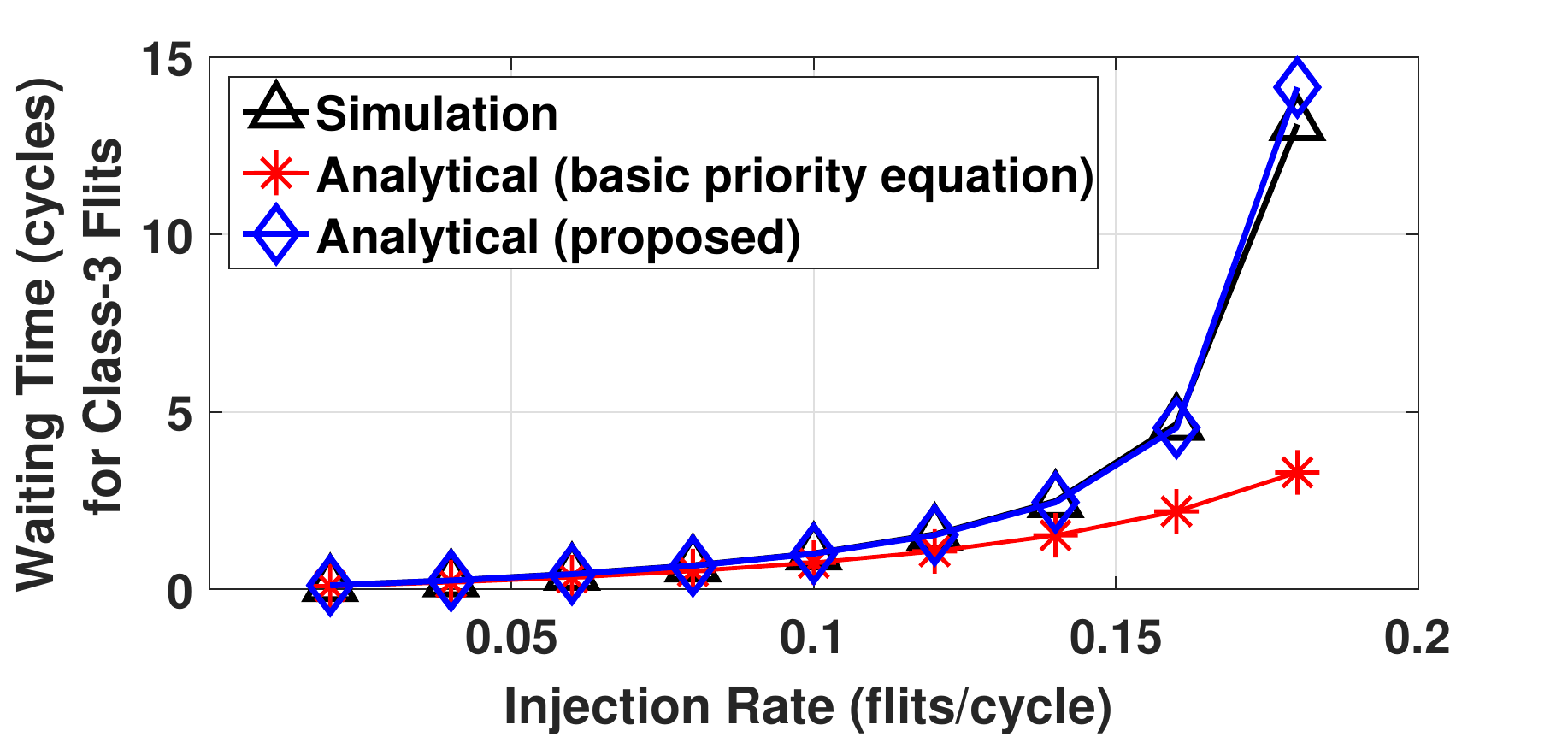}}
	\vspace{-3.5mm}
    \caption{\camready{Comparison of simulation with the basic priority-based queuing model and proposed analytical model.}}
	\vspace{-4.5mm}
	\label{fig:sp_lp_basic_comp}	
\end{figure}
Figure~\ref{fig:sp_hp_basic_comp} shows that the waiting time calculated by the proposed analytical model for flits of traffic class-3 is quite accurate with respect to the waiting time obtained from the simulation.
The average error in waiting time of traffic class-3 is 2\% for the system shown in Figure~\ref{fig_strctr_transform}(a), with a deterministic service time of two cycles.

\subsection{Transformation 2: Split at \textit{Low} Priority Queue} \label{sec:sp_lp}
\noindent \textbf{Conceptual Illustration:}
Consider the queuing system shown in Figure~\ref{fig:servicerate_transform}(a).
In this system, class-1 flits \camready{($\lambda_1$) are waiting in Q$_{high}$,}
while class-2 flits ($\lambda_2$) and class-3 flits ($\lambda_3$) are waiting in Q$_{low}$.
Class-1 and class-3 flits share the same channel and compete for the same output,
\camready{while class-2 flits are sent to a separate output.}
Class-1 flits always win the arbitration since they have higher priority. 
Similar to the previous transformation, the queuing model in Figure~\ref{fig:servicerate_transform}(b) 
is used as an intermediate representation to facilitate the discussion. 
In this system, Q$_{high}$ and Q$_{low}$ are represented as Q$_1$ and Q$_2$ respectively.

If we ignore the impact of class-1 traffic while modeling the waiting time for class-3, 
the resulting analytical models will be highly optimistic, as shown in Figure~\ref{fig:sp_lp_basic_comp}. 
Accounting for the impact of class-1 traffic on class-2 is challenging, since only fraction of the flits in Q$_2$ that compete with class-1 are blocked. 
In other words, class-2 flits which go to the local node are not directly blocked by class-1 flits. 
Hence, there is a need for a new transformation that can address the split at the low-priority queue.


\vspace{1mm}
\noindent \textbf{Proposed Transformation:}
The high-priority flow (class-1) is not affected by \camready{class-2 traffic since they do not share the same server.}
Therefore, the waiting time of class-1 flits can be readily obtained using Equation~\ref{eq:W_basic_prior} as:
\begin{equation} \label{eq:hp_waiting}
W_1 = \frac{R_1 + R_3} {1 - \rho_1}
\end{equation}
Hence, we represent Q$_1$ as a stand-alone queue, as shown in Figure~\ref{fig:servicerate_transform}(c).
However, the opposite is not true; 
class-1 flits affect both class-2 (indirectly) and class-3 (directly). 
Therefore, we represent them using a new queue with modified service rate statistics. 
To ensure that Figure~\ref{fig:servicerate_transform}(c) closely approximates the original system, 
we characterize the effect on the service rate \skm{of class-3} using a novel analytical model.

\noindent \textbf{Proposed Analytical Model:}
Both the service time and residual time of class-3 change due to the interaction with class-1. 
To quantify these changes, we set $\lambda_2 = 0$ such that the effect of class-2 is isolated. 
In this case, the waiting time of class-3 flits can be found using Equation~\ref{eq:W_basic_prior} as:
\begin{equation} \label{eq:W3_at_lambda2_0}
W_{3} \Bigr|_{\lambda_2=0} = \frac{R_1 + R_3 + \rho_1 + \rho_1 W_1} {1 - \rho_1 - \rho_3} 
\end{equation}
We can find $W_3$ also by using the modified service time ($T_3^*$) and 
residual time $R_3^*$ of class-3. 
\camready{The probability that} a class-3 flit cannot be served due to class-1 is equal to server utilization $\rho_{1}$. 
Moreover, there will be extra utilization due to the residual effect of class-3 on class-1, i.e.,
$\lambda_{1} R_{3}$ flits in Q$_1$. 
Hence, the probability that a class-3 flit is delayed due to class-1 flits is:
\begin{equation} \label{eq_p}
p = \rho_{1} + \lambda_{1} R_{3} 
\end{equation}
Each time class-3 flit is blocked by the class-1 flits, the extra delay will be $T_1$, i.e., class-1 service time. Since each flit can be blocked multiple consecutive times, the additional busy period of \camready{serving} class-3 $(\Delta{T}_3)$ \skm{is expressed} as:
\begin{equation} \label{eq_delta_t}
\begin{split}
\Delta{T}_3 &= T_1 p(1-p) + 2T_1 p^2(1-p) + .... + n T_1 p^n(1-p) + \cdots\\
&= T_1\frac{p}{1-p} 
\end{split}
\end{equation}
Consequently, the modified service time ($T_{3}^*$) and utilization ($\rho_{3}^*$) 
of class-3 can be expressed as: 
\begin{align} 
T_{3}^* = T_3 + \Delta T_3 \nonumber \\
\rho_{3}^* = \lambda_{3} T_{3}^* \label{eq_modifed rho}
\end{align}
\begin{algorithm}[b]
\caption{\camready{End-to-end queuing time calculation for different traffic classes}}
\textbf{Input:} Injection rates for all traffic classes, NoC topology and Traffic routing pattern \\
\textbf{Output:} Queuing time for all traffic classes \\
\SetAlgoLined
 \For {n = 1: no. of queues}  { 
 For queuing time expression of the current queue:\\ Initialize: $num_1$ \textleftarrow 0, $den_1$ \textleftarrow 1 \\
 Get all classes in current queue \\ 
 \For {i = 1: no. of classes}{ 	
 Get all higher priority classes than current class \\ 
 For reference queuing time ($W_{ref}$) of current class:
 \\ Initialize: $num_2$ \textleftarrow $R_i$, $den_2$ \textleftarrow (1- $\rho_{i}$)\\ 
 \For {j = 1:no. of higher priority classes}{
 
Calculate coefficient of variation ($C_D$) for current high priority class \\ 
Calculate queuing time expression ($W_{ij}$) and residual time expression ($R_{ij}$) using $C_D$ using \skm{Eq.~\ref{eq:R_1_Q2}, Eq.~\ref{eq:W_1_Q2}, and Eq.~\ref{eq:W_2_Q2}}\\
 $num_2$ \textleftarrow $num_2 + R_{ij} + \rho_{ij} + W_{ij} \rho_{ij}$ \\
 $den_2$ \textleftarrow $den_2 - \rho_{ij}$
}
 $W_{ref} = \frac{num_2}{den_2}$ \skm{(Eq.~\ref{eq:W3_at_lambda2_0})} \\
Modify service rate ($T_{i}^*$) of $i^{\mathrm{th}}$class \\ 
Calculate residual time ($R_{i}^*$) using $T_{i}^*$ and $W_{ref}$ using \skm{Eq.~\ref{eq_R_modified}}\\
 $num_1$ \textleftarrow $num_1 + R_{i}^*$ \\ 
 $den_1$ \textleftarrow $den_1$ - $\rho_i^*$
}
\skm{Queuing time of class-i in $n^{\mathrm{th}}$ queue = $\frac{num_1}{den_1} + \Delta T_i$}
}
\label {algo_model_generation}
\vspace*{-1mm}
\end{algorithm}
Suppose that the modified residual time of class-3 is denoted by $R_{3}^*$. 
We can plug $R_{3}^*$, the modified utilization $\rho_{3}^*$ from Equation~\ref{eq_modifed rho}, 
and the additional busy period $\Delta{T}_3$ from Equation~\ref{eq_delta_t} into Geo/G/1 model
to express the waiting time $W_3$ as:
\begin{equation} \label{eq_W_ref}
W_3 = \frac{R_{3}^*}{1 - \rho_{3}^*} + \Delta{T}_3
\end{equation}
When $\lambda_2$ is set to zero, this expression should give the class-3 waiting time 
$W_{3} \Bigr|_{\lambda_2=0}$ found in Equation~\ref{eq:W3_at_lambda2_0}. 
Hence, we can find the following expression for $R_{3}^*$ by 
\camready{combining} 
Equation~\ref{eq:W3_at_lambda2_0} and Equation~\ref{eq_W_ref}:
\begin{align} \label{eq_R_modified}
 R_{3}^* = (1 - \rho_{3}^*) (W_{3} \Bigr|_{\lambda_2=0} - \Delta{T}_3)
\end{align}
Since the modified service time and residual times are computed, 
we can apply the Geo/G/1 queuing model one more time to find the waiting time of class-2 and class-3 
flits as:
\begin{align} \label{eq_W2}
W_{2} &= \frac{R_{3}^* + R_{2}} {1 - \rho_{3}^* - \rho_2 } \nonumber \\
%
%
W_{3} &= \frac{R_{3}^* + R_{2}} {1 - \rho_{3}^* - \rho_2 } + \Delta T_{3}
\end{align}
%
%
%
%
%
Figure \ref{fig:sp_lp_basic_comp} shows that the class-3 waiting time calculated using the proposed analytical modeling technique is very close to simulation results. The modeling error is within 4\% using a deterministic service time of 2 cycles.

\section{Generalization for Arbitrary Number of Queues} \label{sec:model_gen}
In this section, we show how the proposed transformations are used to generate analytical models for priority-based NoCs with arbitrary topologies and input traffic. 
\camready{Algorithm~1 describes the model generation technique, which is a part of the proposed methodology to be used in a virtual platform.}
This algorithm takes injection rates for all traffic classes, the NoC topology, and the routing \camready{of individual traffic classes}. 
Then, it uses the transformations described in Section \ref{sec:sp_hp} and Section \ref{sec:sp_lp} iteratively to construct analytical performance models for each traffic class.

\begin{figure*}[b]
	\centering
	\includegraphics[width=1\textwidth]{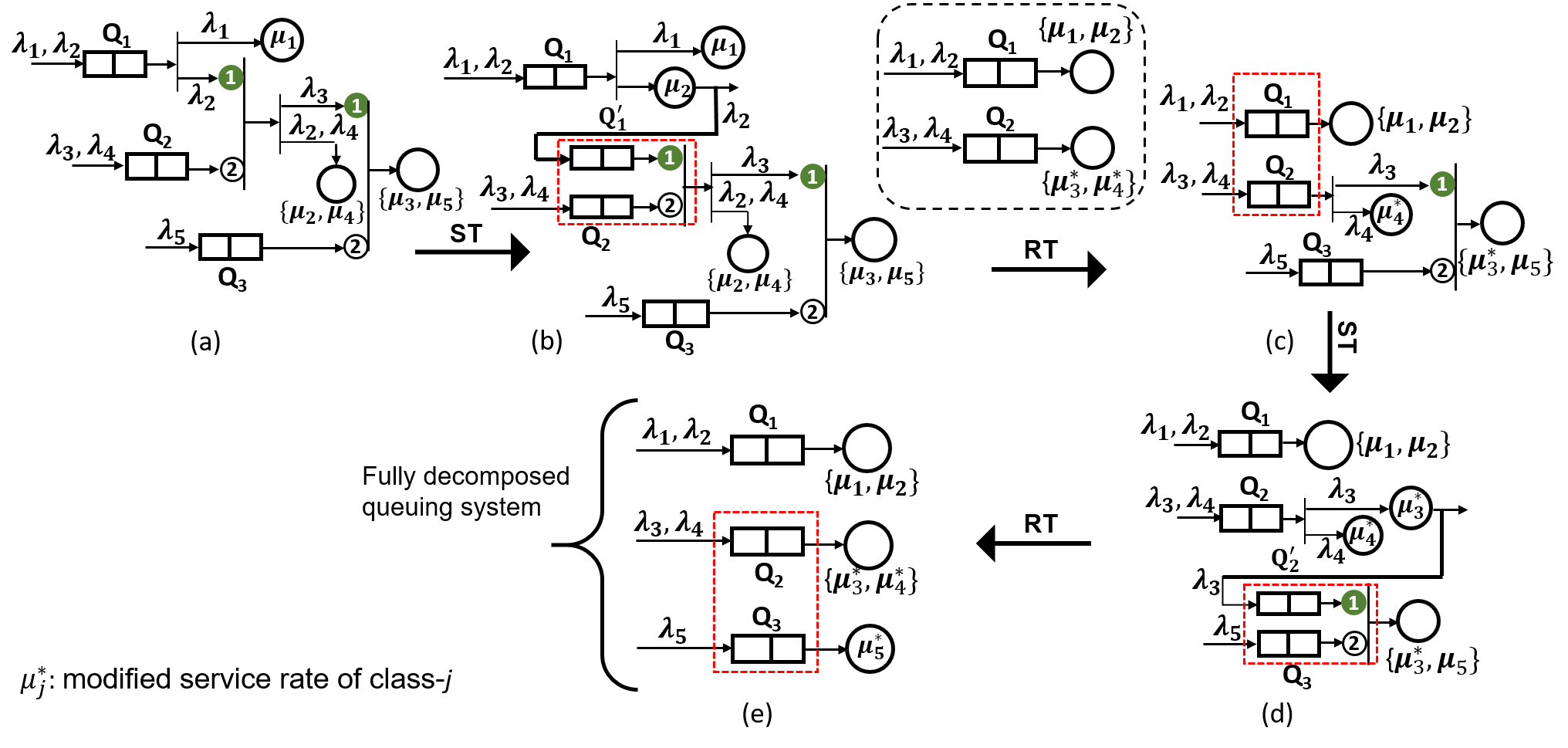}
	\vspace{-7mm}
	\caption{\revfinal{Applying the proposed methodology on a representative segment of a priority-based network. ST and RT denote Structural and Service Rate Transformation, respectively. Red-dotted squares show the transformed part from the previous step. Figure (a) shows the original queuing system. After applying ST on Q$_1$, we obtain the system shown in Figure (b). The system in Figure (c) is obtained by applying RT on Q$_2$. ST is applied again on Q$_2$ to obtain the system shown in Figure (d). Finally, RT is applied on Q$_3$ to obtain the fully decomposed queuing system shown in Figure (e).}}
    \vspace{-1mm}
	\label{fig:compl_example}	
\end{figure*}
First, Algorithm~1 extracts all traffic classes originating from a particular queue, as shown in line 6.
Next, the waiting time for each of these classes is computed separately, as each has a different dependency on other classes due to priority arbitration. 
At line 8, all classes that have higher priority than the current class are obtained.
In lines 11--16, the structural transformation as described in Section \ref{sec:sp_hp} is applied.
For that, the coefficient of variation of inter-departure time $(C_D)$ for each of the higher priority classes is computed.
Through structural transformation, reference waiting time $(W_{ref})$ for the current class is obtained, as depicted in line 17 of the algorithm.
At line 18, we compute the modified service time $(T_i^*)$ of the current class following the method described in Section~\ref{sec:sp_lp}. 
Using $T_i^*$ and $W_{ref}$, the residual time ($R_i^*$ in line 19) is computed. 
Using residual time expressions for all classes in a queue, we obtain waiting time expressions for each class separately, as shown in line 23 of the algorithm.

%
%
Figure~\ref{fig:compl_example} illustrates the proposed approach on a representative example of a priority-based network to decompose the system.
Figure \ref{fig:compl_example}(a) shows the original queuing network.
This network consists of three queues: Q$_1$, Q$_2$, and Q$_3$.
Q$_1$ stores flits from class-1 and class-2 flows, 
while Q$_2$ buffers class-3 and class-4 flits. 
Flits of class-2 have higher priority than both class-3 and class-4, 
as denoted by the first port of the switch that connects these flows. 
Finally, class-5 flits are stored in Q$_3$.
We note that class-5 flits have lower priority than that of class-3, 
while they are independent of class-2 and class-4 flits.
To solve this queuing system, we first apply the structural transformation on class-1 and class-2 of Q$_1$ by bypassing class-2 flits to Q${_1\textprime}$ as shown in Figure~\ref{fig:compl_example}(b).
Next, the service rate transformation on class-3 and class-4 is applied to obtain modified service time $(\mu^*)$.
This transformation allows us to form the network by decomposing Q$_1$ and Q$_2$, 
as depicted in Figure~\ref{fig:compl_example}(c).
After that, structural transformation is applied on class-3 as flits of class-3 have higher priority than those of class-5.
Finally, service rate transformation is performed on class-5 to achieve a fully decomposed system, which is shown in Figure~\ref{fig:compl_example}(e).
\vspace{1mm}
\noindent \textbf{Automation of Model Generation Technique:} 
We developed a framework to automatically generate the analytical performance model for NoCs with arbitrary size 2D Mesh and ring topologies.
The proposed framework operates in two steps.
In the first step, we extract all architecture-related information of the NoC.
This includes information about the traffic classes in each queue and priority relations between classes.
In the second step, the automation framework uses this architecture information to generate analytical models.

\section{Experimental evaluations} \label {sec:exp_eval}

\subsection{Experimental Setup}

We applied the proposed analytical models to a widely used priority-based industrial NoC design~\cite{jeffers2016intel}. 
\revfinal{We implemented the proposed analytical models in C and observed that on average it takes $0.66\mu s$ to calculate latency value per source-to-destination pair.}
At each router of the NoC, there are queues in which tokens wait to be routed.
This NoC design incorporates deterministic service time across all queues.
We compared average latency values in the steady state found in this approach against an industrial cycle-accurate simulator written in SystemC~\cite{Xplore,ogras2012energy}.
We ran each simulation for 10 million cycles to obtain steady state latency values, with a warm-up period of 5000 cycles.
Average latency values are obtained by averaging latencies of all flits injected after the warm-up period. 
Injection rates are swept from $\lambda_1$ to $\lambda_{max}$. Beyond $\lambda_{max}$, server utilization becomes greater than one, which is not practical.
We show the average latency of flits as a function of the flit injection rate for different NoC topologies.
\revfinal{We also present experimental results considering the cache coherency protocol with different hit rates, network topologies, and floorplans. With a decreasing hit rate, traffic towards the memory controller increases, leading to more congestion in the network.}

%
\subsection{Full-System Simulations on gem5}

\begin{figure}[h]
	\centering
	\includegraphics[width=0.6\columnwidth]{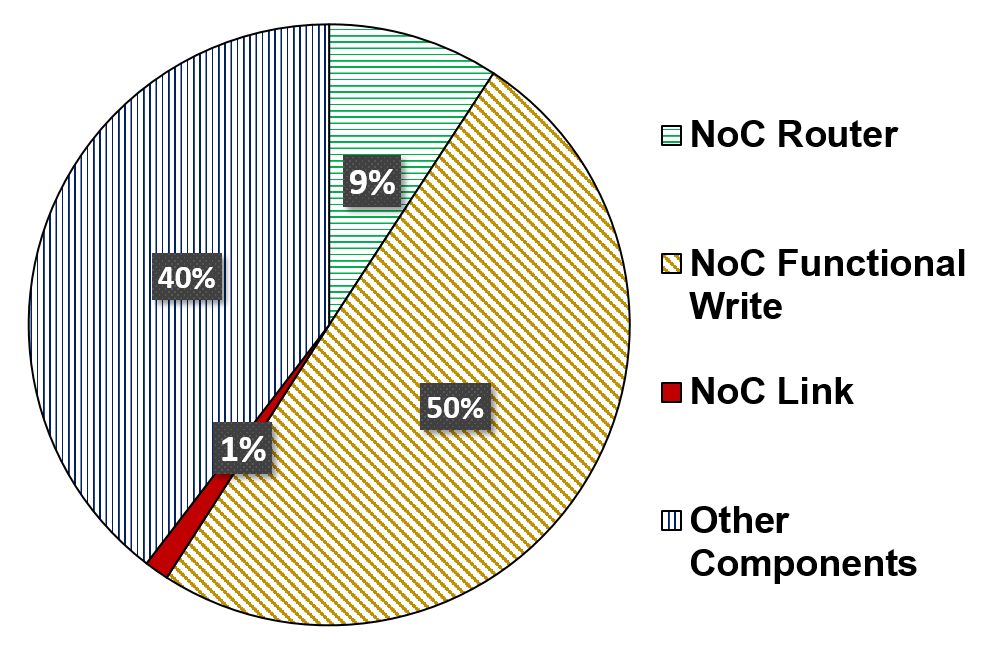}
	\vspace{-0mm}
	\caption{The fraction of simulation time spent by different functions while running Streamcluster in gem5. NoC-related functions take 60\% of simulation time.} 
	\label{fig:app_profile}
\end{figure}

Applications are profiled in the full-system simulator gem5~\cite{Binkert2011Gem5} using Linux `perf' tools~\cite{de2010new}.
The `perf' tool captures the time taken by each function call and their children in the gem5 source.
It represents the statistics through a function call graph.
From this call graph, we obtain the time taken by the functions related to Garnet2.0, which is the on-chip interconnect for gem5.
Figure~\ref{fig:app_profile} shows components of Garnet2.0, which takes up a significant portion of the total simulation time while running Streamcluster application on gem5.
These components are router, network-link, and functional write.
The `other components' shown in Figure~\ref{fig:app_profile} consists of the functions not related to network simulation.
We observe that the functional write takes 50\%, and the whole network takes around 60\% of the total simulation time in this case.

\noindent \revfinal{\textbf{Simulation Time:} 
To evaluate the decrease in simulation time with the proposed approach, 
we first run the Streamcluster application with a 16-core CPU on gem5 in full system mode using Garnet2.0, a cycle-accurate network simulator. 
Then, we repeat the same simulation by replacing the cycle-accurate simulation with the proposed analytical model. 
The total simulation time is reduced from 12,466 seconds to 4986 seconds
when we replace the cycle-accurate NoC simulations with the proposed analytical models.
Hence, we achieve a 2.5$\times$ speedup in cycle-accurate full-system simulation with the proposed NoC performance analysis technique.}


\subsection{Validation on Ring Architectures} \label{Sec:half_ring_result}
\begin{figure}[h]
	\centering
	\resizebox{0.6\columnwidth}{!}{\includegraphics[width=0.5\textwidth]{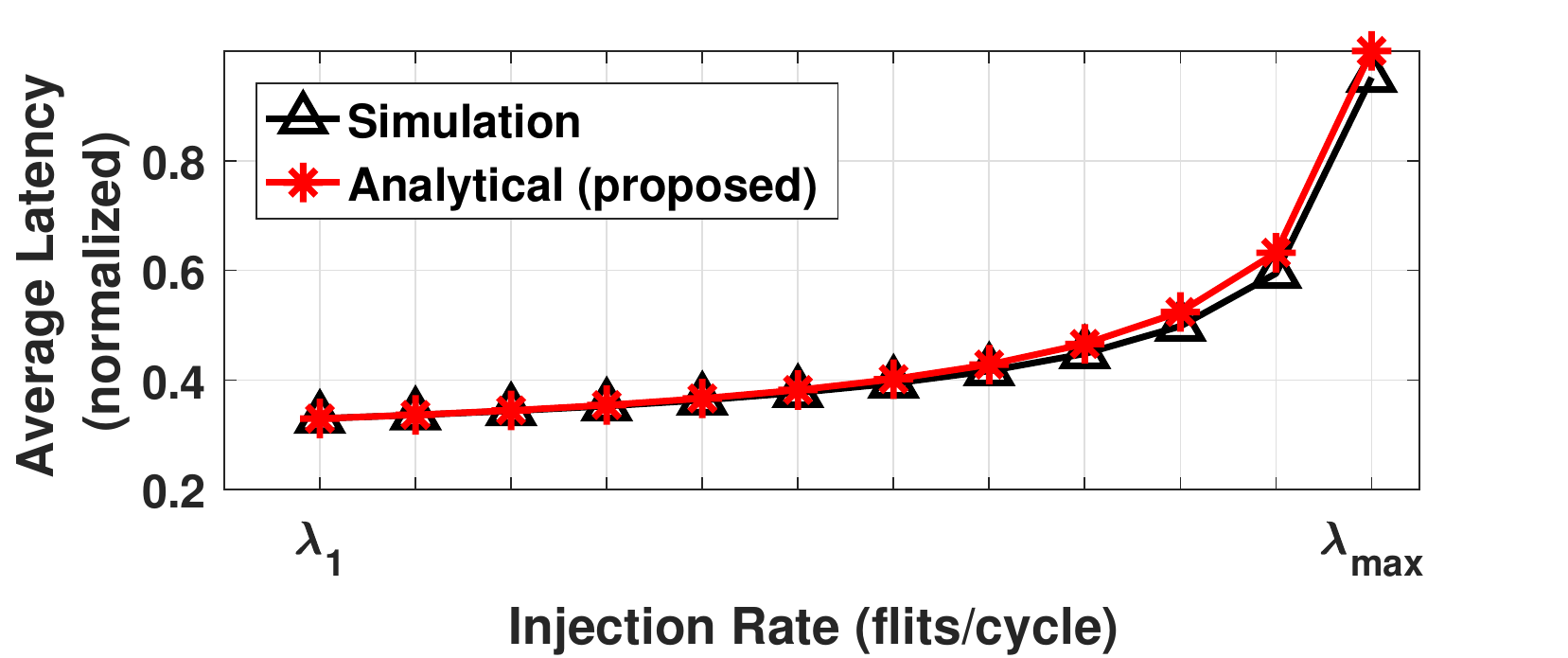}}
    \vspace{-3mm}
	\caption{Evaluation of the proposed model on a ring with eight nodes.}
	\label{fig:8_ring}	
\end{figure}

This section evaluates the proposed analytical models on priority-based ring architecture that consists of eight nodes.
In this experiment, all nodes inject flits with an equal injection rate.
Flits injected from a node go to other nodes with equal probability.
We obtain the latency between each source-destination pair using the proposed analytical models.
\revfinal{The simulation and analysis results are compared in Figure \ref{fig:8_ring}. 
The proposed analysis technique has only 2\% error on average.
The accuracy is higher at lower injection rates and degrades gradually with increasing  injection rates, as expected. 
However, the error at the highest injection rate is only 5.2\%.} 

%
\subsection{Validation on Mesh Architectures} \label{sec:mesh_result}
\begin{figure*}[h]
    \begin{minipage}{0.48\textwidth}
	\centering
	\resizebox{1.0\columnwidth}{!}{\includegraphics[width=0.8\textwidth]{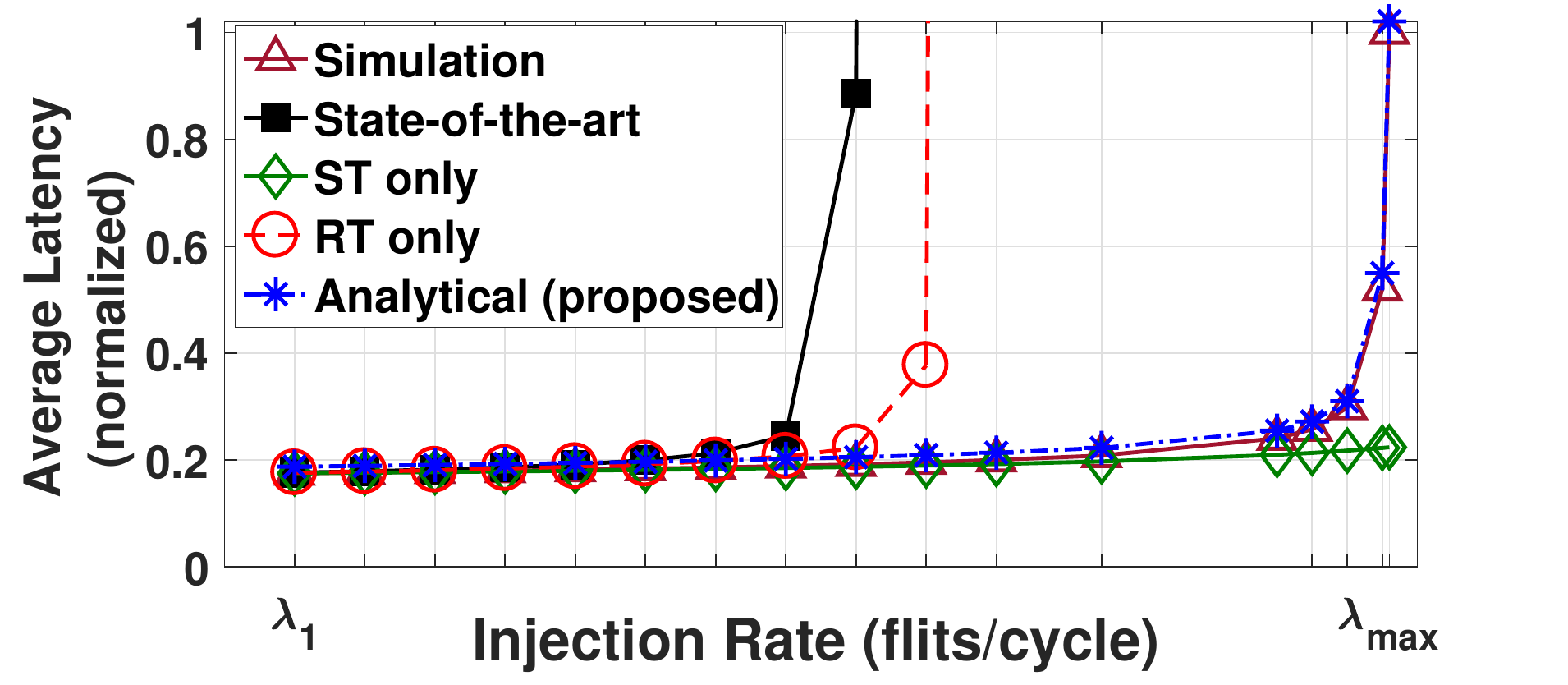}}
	\vspace{-3mm}
	\caption{\revfinal{Evaluation of the proposed model \\ on a 6$\times$6 mesh.}}
	\label{fig:6x6_mesh}
	\end{minipage} \hfill
    \begin{minipage}{0.48\textwidth}
	\centering
	\resizebox{1.0\columnwidth}{!}{\includegraphics[width=0.8\textwidth]{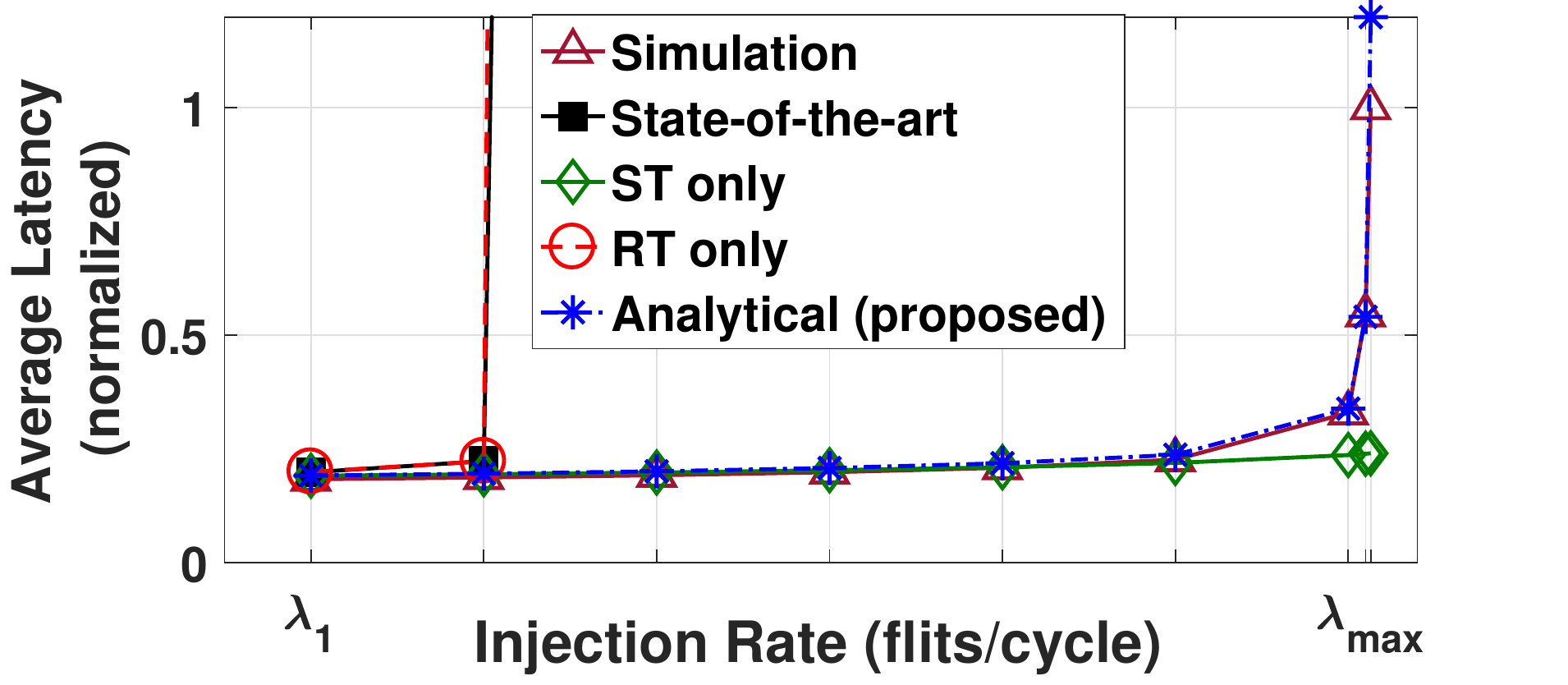}}
	\vspace{-3mm}
	\caption{\revfinal{Evaluation of the proposed model \\ on an 8$\times$8 mesh.}}
	\label{fig:8x8_mesh}
	\end{minipage}
\end{figure*}

This section evaluates the proposed analytical model for 6$\times$6 and 8$\times$8
priority-based mesh NoCs with Y-X routing. 
As described in \cite{jeffers2016intel}, a mesh is a combination of horizontal and vertical half rings. 
The analytical model generation technique for priority-based NoC architecture is applied to horizontal and vertical rings individually. 
Then, these latencies, as well as the time it takes to switch from one to the other are used to obtain the latency for each source-destination pair.
We first consider uniform random all-to-all traffic, as in Section \ref{Sec:half_ring_result}.
The comparison with the cycle-accurate simulator shows that the proposed analytical models are on average 97\% and 96\% accurate for 6$\times$6 and 8$\times$8 mesh, 
as shown in Figure~\ref{fig:6x6_mesh} and Figure~\ref{fig:8x8_mesh}, respectively.
At the highest injection rate, the analytical models show 11\% error for both cases.

\noindent \revfinal{\textbf{Comparison to Prior Techniques:}
We compare the proposed analytical models to the existing priority-aware analytical models in literature~\cite{walraevens2004discrete}. 
Since these techniques do not consider multiple priority traffic classes in the network, they fail to accurately estimate the end-to-end latency. For example, Figure~\ref{fig:6x6_mesh} and Figure~\ref{fig:8x8_mesh} show that they overestimate NoC latency at high injection rates for 6$\times$6 and 8$\times$8 mesh networks, respectively. 
In contrast, since it captures the interactions between different classes, the proposed technique is able to estimate latencies accurately.
Finally, we analyze the impact of using each transformation individually. 
If we apply only the Structural Transformation (ST), 
then the latency is severely underestimated at higher injection rates, 
since contentions are not captured accurately. 
In contrast, applying only Service Rate Transformation (RT) results in overestimating the latency at higher injection rates as
the model becomes pessimistic.}

\begin{figure}[h]
	\centering
	\resizebox{0.6\columnwidth}{!}{\includegraphics[width=0.5\textwidth]{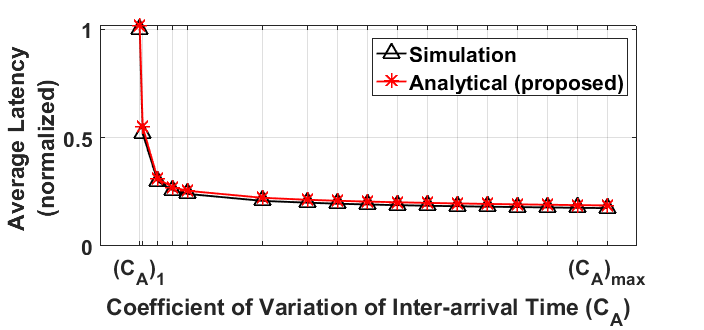}}
	\vspace{-3mm}
	\caption{\revfinal{Effect of coefficient of variation of inter-arrival time on average latency for a 6$\times$6 mesh.}}
	\label{fig:6x6_mesh_ca}	
\end{figure}

\vspace{1mm}
\noindent\revfinal{\textbf{Impact of coefficient of variation:} One of the important parameters in our analytical model is the coefficient of variation of inter-arrival time. When the inter-arrival time between the incoming flits follows geometric distribution, increasing coefficient of variation implies larger inter-arrival time. Hence, the average flit latency is expected to decrease with an increasing coefficient of variation. Indeed, the simulation and analysis results demonstrate this behavior for a 6$\times$6 mesh in Figure~\ref{fig:6x6_mesh_ca}. We observe that the proposed technique accurately estimates the average latency in comparison to cycle-accurate simulation. On average, the analytical models are 97\% accurate with respect to latency obtained from the simulation in this case.}

\vspace{1mm}

\noindent\textbf{\camready{{Evaluation with Intel\textsuperscript{\textregistered} Xeon\textsuperscript{\textregistered} Scalable Server Processor Architecture:}}}
This section evaluates the proposed analytical model with the floorplan of a variant of the \camready{Intel\textsuperscript{\textregistered} Xeon\textsuperscript{\textregistered} Scalable Server Processor Architecture~\cite{doweck2017inside} architecture.}
This version of the Xeon server has 26 cores, 26 banks of the last level cache (LLC), and 2 memory controllers.
\begin{figure}[t]
		\centering
		\resizebox{0.6\columnwidth}{!}{\includegraphics[width=0.5\textwidth]{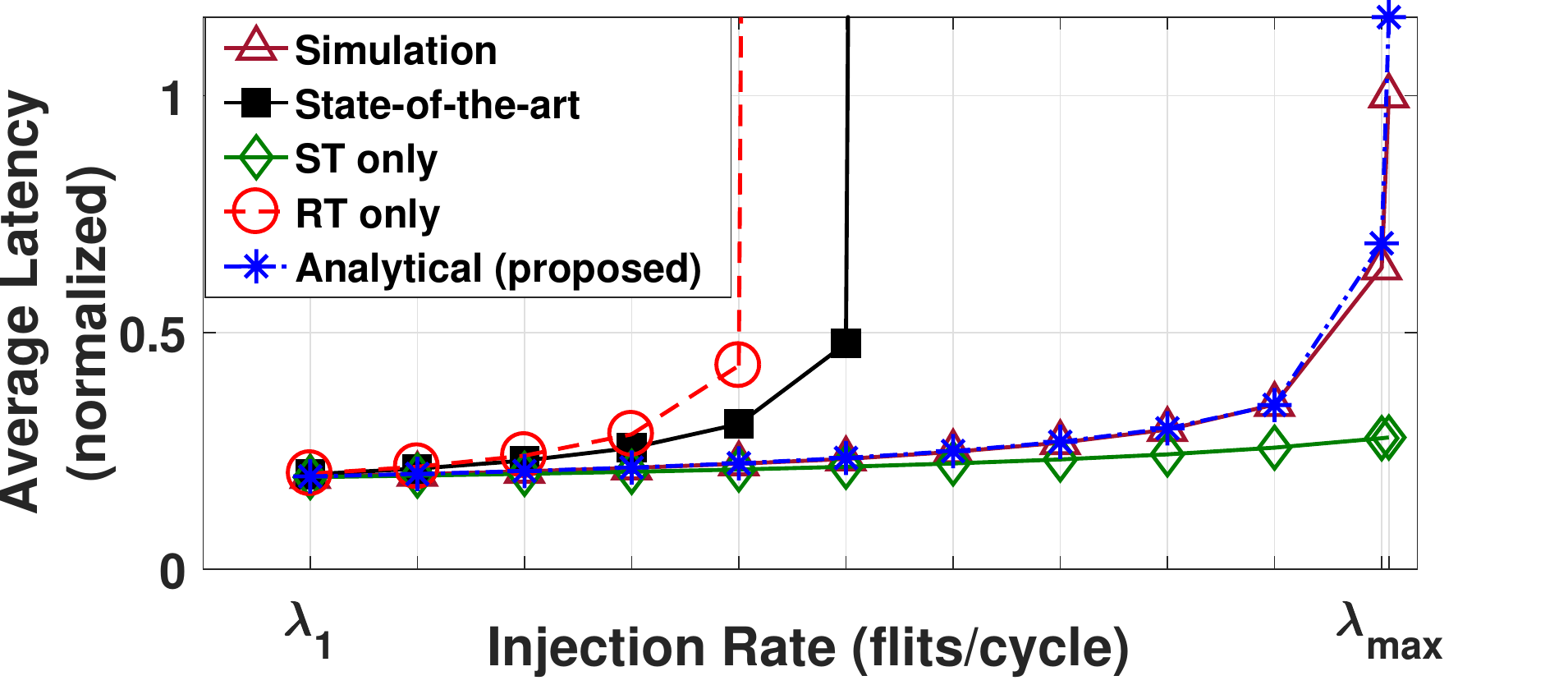}}
		\vspace{-3mm}
		\caption{\revfinal{Evaluation of the proposed model on one variant of the Xeon server architecture.}}
		\label{fig:floorplan}	
\end{figure}
The cores and LLC are distributed on a 6$\times$6 mesh NoC.
The comparison of simulation and proposed analytical models with this floorplan is shown in Figure~\ref{fig:floorplan}.
On average, the accuracy is 98\% when all cores send flits to all caches with equal injection rates.
Similar to the evaluations on 6$\times$6 mesh and 8$\times$8 mesh, the state-of-the-art NoC performance analysis technique~\cite{walraevens2004discrete} highly overestimates the average latency for this server architecture, as shown in Figure ~\ref{fig:floorplan}.
Applying only ST underestimates the average latency and applying only RT overestimates the average latency.

\begin{figure}[h]
		\centering
		\resizebox{0.6\columnwidth}{!}{\includegraphics[width=0.5\textwidth]{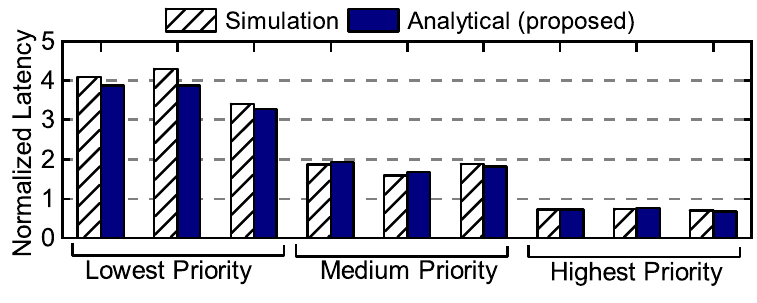}}
		\vspace{-3mm}
		\caption{\revfinal{Per-class latency comparison for the server example.}}
		\label{fig:sim_ana_per_class}	
\end{figure}

The NoC latency is a function of the traffic class, since higher priority classes experience less contention. To demonstrate the latency for different classes, we present the NoC latencies for 9 representative traffic classes of the server architecture described above. Figure~\ref{fig:sim_ana_per_class} shows the latency of each class \camready{of the server architecture described above} normalized with respect to the average latency obtained from the simulation. Higher priority classes experience lower latency, as expected. The proposed performance analysis technique achieves 91\% accuracy on average for the classes which have the lowest priority in the NoC. For the classes having medium priority and highest priority, the accuracy is 99\% on average. Therefore, the proposed technique is reliable for all classes with different levels of priority.

\begin{table}[t]
\caption{Accuracy for cache-coherency traffic flow}
\vspace{-3mm}
\label{tab:cache_coher}
\begin{tabular}{|c|c|c|}
\hline
\begin{tabular}[c]{@{}c@{}}LLC\\ Hit Rate (\%)\end{tabular} & \begin{tabular}[c]{@{}c@{}}Accuracy for\\ Address Network (\%)\end{tabular} & \begin{tabular}[c]{@{}c@{}}Accuracy for\\ Data Network (\%)\end{tabular} \\ \hline
100                                 & 98.8                                                                     & 93.9                                                                  \\ \hline
50                                  & 97.7                                                                     & 98.1                                                                  \\ \hline
0                                   & 97.7                                                                     & 98.0                                                                  \\ \hline
\end{tabular}
\vspace{-3mm}
\end{table}
%


Finally, we evaluate the proposed technique with different LLC hit rates. Table~\ref{tab:cache_coher} shows that the proposed approach achieves over 97\% accuracy in estimating the average latency of the address network for all hit rates. 
Similarly, the latencies in the data network are estimated with 98\% \camready{or greater} accuracy for 0\% and 50\% hit rates. 
The accuracy drops to 93.9\% for 100\% hit rates, since this scenario leads to the highest level of congestion due to all-to-all traffic behavior.


\subsection{Evaluation with Real Applications} \label{sec:real_apps}
In this section, evaluations of the proposed technique with real applications are shown.
We use gem5~\cite{Binkert2011Gem5} to extract traces of applications in Full-System (FS) mode.
Garnet2.0~\cite{agarwal2009garnet} is used as the network simulator in gem5 with the Ruby memory system.
Table~\ref{tab:gem5_config} shows the various configuration settings we used for FS simulation in gem5.

\begin{table}[b]
\caption{Configuration settings in the gem5 simulation}
\vspace{-3mm}
\label{tab:gem5_config}
\begin{tabular}{l|ll}
\hline
\multirow{3}{*}{\textbf{Processor}}                                                      & Number of Cores    & 16                                                                                             \\ \cline{2-3} 
                                                                                         & Frequency of Cores & 2 GHz                                                                                          \\ \cline{2-3} 
                                                                                         & Instruction Set    & x86                                                                                            \\ \hline
\multirow{2}{*}{\textbf{\begin{tabular}[c]{@{}l@{}}Interconnect\\ Network\end{tabular}}} & Topology           & 4x4 Mesh                                                                                       \\ \cline{2-3} 
                                                                                         & Routing Algorithm  & X-Y deterministic                                                                              \\ \hline
\multirow{2}{*}{\textbf{\begin{tabular}[c]{@{}l@{}}Memory\\ System\end{tabular}}}        & L1 Cache           & \begin{tabular}[c]{@{}l@{}}16KB of instruction\\  and data cache \\ for each core\end{tabular} \\ \cline{2-3} 
                                                                                         & Memory Size        & 3 GB                                                                                           \\ \hline
\multirow{2}{*}{\textbf{Kernel}}                                                         & Type               & Linux                                                                                          \\ \cline{2-3} 
                                                                                         & Version            & 3.4.112                                                                                        \\ \hline
\end{tabular}
\end{table}

We collect traces of six 16-threaded applications from PARSEC~\cite{bienia2008parsec} benchmark suites: Blackscholes, Canneal, Swaptions, Bodytrack, Fluidanimate, and Streamcluster. 
We selected applications that show relatively higher network utilization as discussed in~\cite{wettin2014performance}. 
The accuracy obtained for these applications is an important indicator of the practicality of the proposed technique since real applications do not necessarily comply with a known inter-arrival time distribution~\cite{bogdan2011non}, such as the geometric distribution used in this work.
The traces are parsed and simulated through our custom in-house simulator with priority-based router model.
For each application, a window of one million cycles with the highest injection rate is chosen for simulation.
From the traces of these applications, we get the average injection rate of each source and destination pair.
These injection rates are fed to our analytical models to obtain average latency.

\begin{figure*}[t]
		\centering
		\includegraphics[width=1\textwidth]{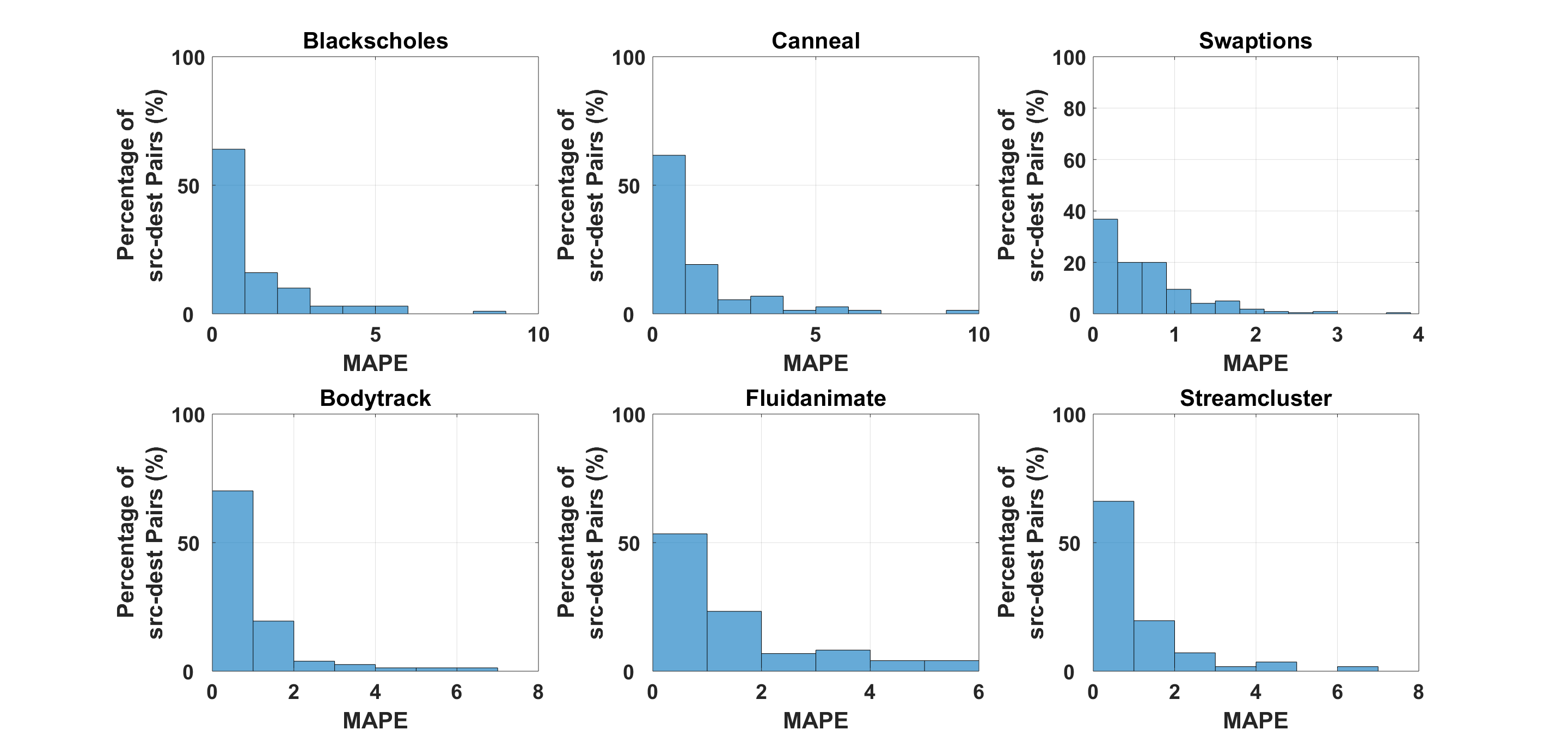}
		\vspace{-2mm}
		\caption{Model comparison for different applications from PARSEC suite.}
		\label{fig:benchmark_compare}	
\end{figure*}

Figure~\ref{fig:benchmark_compare} shows the comparison of the average latency between the proposed analytical model and the simulation.
The x-axes represent mean absolute percentage error (MAPE) between the average simulation latency ($L_{sim}$) and average latency obtained from analytical models ($L_{analytical}$).
MAPE is defined by the following equation:
\begin{equation}
    \text{MAPE} = 100 \bigg( \frac{|L_{sim} - L_{analytical}|}{L_{sim}} \bigg)
\end{equation}
The y-axes in the plots represent the percentage of source to destination pairs having the corresponding MAPE.
From this figure, we observe that the latency obtained from the proposed analytical model is always within 10\% of the latency reported by the cycle-accurate simulations.
In particular, \camready{only 1\% source-destination pair has MAPE of 10\% for the Canneal application.}
On average, the analytical models have 3\% error in comparison to latency obtained from the simulation for real applications.
\revfinal{These results demonstrate that our technique achieves high accuracy for applications which may have arbitrary inter-arrival time distributions.}

\begin{figure}[b]
		\centering
		\includegraphics[width=0.6\textwidth]{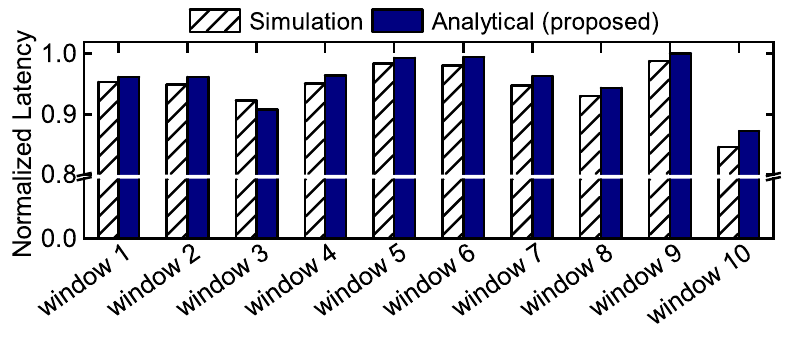}
		\caption{\revfinal{Evaluation of the proposed model under a finer level of time granularity (100K cycles) for Streamcluster application.}}
		\label{fig:latency_window}	
\end{figure}

\revfinal{We further divide the window of one million cycles into 10 smaller windows containing 100,000 cycles each. Average latency comparison for Streamcluster application in these smaller windows is shown in Figure~\ref{fig:latency_window}. 
The largest MAPE between latency obtained from the simulation and analytical model is observed for window 10, which is 7\%. 
On average, the proposed analytical models are 98\% accurate for these 10 windows.
This confirms the reliability of the proposed analytical models at an even more granular level for the application.}
Finally, we note that the experiments with synthetic traffic shown in Section~\ref{Sec:half_ring_result} and Section~\ref{sec:mesh_result} exercise \skm{higher injection rates} than these applications.
Hence, the proposed technique performs well both under real application traces and heavy traffic scenarios.

\revfinal{Prior work showed that the deviation from Poisson distribution becomes larger as the network load approaches saturation~\cite{ogras2010analytical}. Similar to this result, we also observe that the Geometric distribution assumption is a good approximation until the NoC operates near saturation point. Therefore, we obtain high accuracy for real application workloads. Since this accuracy can degrade with increasing traffic load, we plan to generalize the proposed models by relaxing the assumption of Geometric distribution in our future work.}


\section{Conclusion} \label{sec:concl_future}
\vspace{2mm}
In this work, we propose an approach to build analytical models for priority-based NoCs with multi-class flits. As we emphasized, no prior work has presented analytical models that consider priority arbitration and multi-class flits in a single queue simultaneously. Such a priority-based queuing network is decomposed into independent queues using novel transformations proposed in this work. We evaluate the efficiency of the proposed approach by computing end-to-end latency of flits in a realistic industrial platform
and using real application benchmarks. 
Our extensive evaluations show that the proposed technique achieves a high accuracy of 97\% accuracy compared to cycle-accurate simulations for different network sizes and traffic flows.
\vspace{3mm}

\let\secfnt\undefined
\bibliographystyle{abbrv}

\section*{Appendix A} \label {sec: residual time}
\begin{figure}[t]
	\centering
	\resizebox{0.7\textwidth}{!}{\includegraphics[width=1\textwidth]{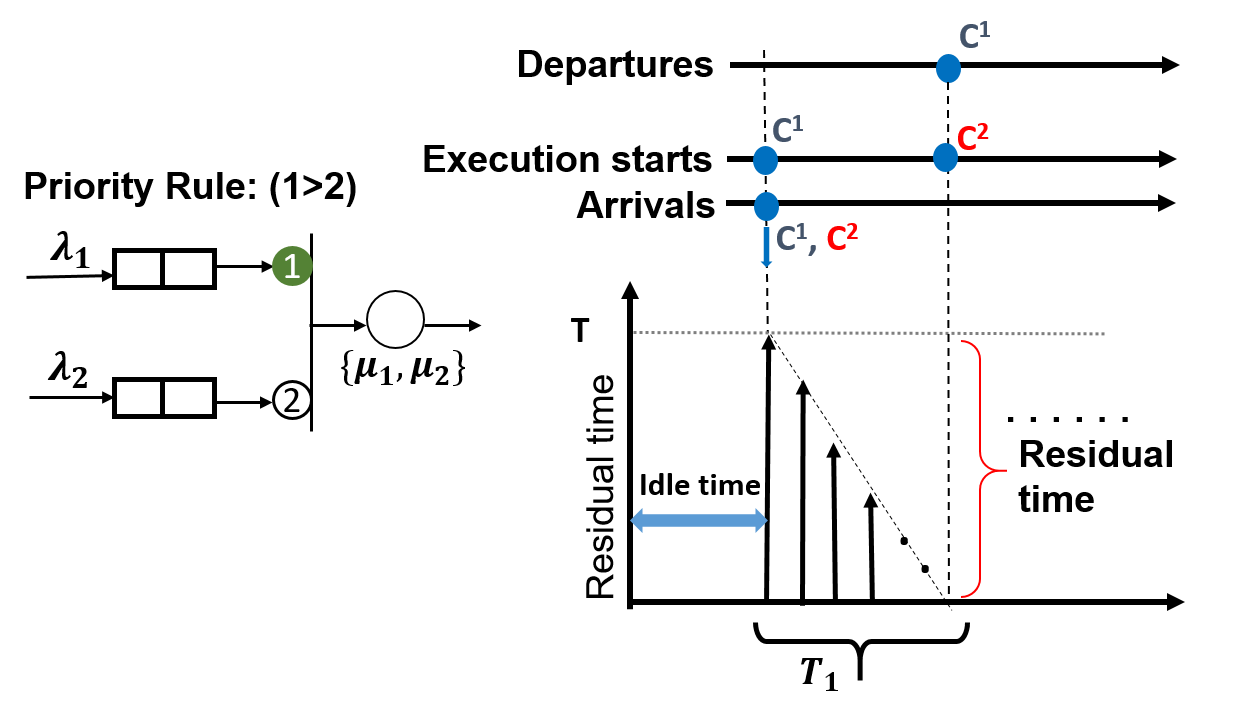}}
	\vspace{-5mm}
	\caption{Residual time calculation.}
	\vspace{-5mm}
	\label{fig_residualtime}
\end{figure}

\textbf{Residual time calculation: }Residual time is the delay of serving the next flit due to the remaining service time for a currently processed flit. As illustrated in Figure \ref{fig_residualtime}, class-2 flits (low-priority flits) have to wait until the server becomes free.
\skm{Equations~\ref{eq_R_1} and \ref{eq_R_2} are expressions for total residual time effect on class-1 and class-2 respectively.}
\camready{The analytical models for residual time have been well studied for continuous systems, yet little attention was given for discrete systems~\cite{bertsekas1992data}.}
\camready{In this section, we derive analytical expressions of residual time for priority-based queuing systems using discrete time domain analysis}.
These equations are derived assuming that the arrival process at each queue follows geometric distribution. 
Average residual time for each class of flit is evaluated by averaging the area of the \camready{residual time} triangles shown in Figure \ref{fig_residualtime} over all flits injected in a large amount of time.
Let us assume that $M_1(T_{tot})$ and $M_2(T_{tot})$ are the total numbers of flits of class-1 and class-2 respectively which are injected into the system in $T_{tot}$ amount of time and $T_i$ is the service time for $i^{\mathrm{th}}$ flit.
When a new service duration of $T_i$ begins, then the residual time of ($T_i - 1$) starts and decays linearly.
If we take time average of the residual time, then we obtain Equation~\ref{eq_R_1}: 
\begin{align}  \label{eq_R_1} \nonumber
R^{(1)} &= \frac{1}{T_{tot}}\sum_{i=1}^{M_1(T_{tot})} \Big( \sum_{\tau=0}^{T_i-1} \tau \Big) + \frac{1}{T_{tot}} \sum_{j=1}^{M_2(T_{tot})}\Big( \sum_{\tau=0}^{T_j-1} \tau\Big) \\ \nonumber
&= \frac{1}{T_{tot}}\sum_{i=1}^{M_1(T_{tot})} \Bigg( \frac {T_i(T_i-1)} {2} \Bigg) + \\ 
& \frac{1}{T_{tot}} \sum_{j=1}^{M_2(T_{tot})} \Bigg( \frac {T_j(T_j-1)} {2} \Bigg)
\end{align}
In the derivation of Equation~\ref{eq_R_1}, $\tau$ \camready{is an auxiliary variable that} represents different residual time values for a particular flit.
Multiplying and dividing the first expression in the summation by $M_1(T_{tot})$ and second expression by $M_2(T_{tot})$, we obtain:

\begin{align} \label{eq_R_1_final} \nonumber
R^{(1)}&= \frac {M_1(T_{tot})} {T_{tot}} \frac {1} {M_1(T_{tot})}\sum_{i=1}^{M_1(T_{tot})} \Bigg( \frac {T_i^2-T_i} {2} \Bigg) + \\ \nonumber
& \frac {M_2(T_{tot})} {T_{tot}} \frac {1} {M_2(T_{tot})}\sum_{j=1}^{M_2(T_{tot})} \Bigg( \frac {T_j^2-T_j} {2} \Bigg) \\ \nonumber
&\stackrel{(a)}{=} \frac {1}{2} \lambda_1 (\overline{T_1^2}-\overline{T_1}) + \frac {1}{2} \lambda_2 (\overline{T_2^2}-\overline{T_2}) \\
&\stackrel{(b)}{=} R_1 + R_2 
\end{align}
Where $(a)$ follows from the fact that $\frac{M_i(T_{tot})}{T_{tot}}$ is the average number of injected flits in one time unit, which is $\lambda_i$.
Also, $\overline{T_i}$ and $\overline{T_i^2}$ denote the first and second order moments of the service time of class-i.
We obtain $(b)$ because $\frac {1}{2} \lambda_i (\overline{T_i^2}-\overline{T_i})$ is the residual time of class-i ($R_i$).

\skm{Similarly, we compute the effective residual time of class-2 ($R^{(2)}$).}
At any cycle, both class-1 and class-2 flits can arrive in the system.
If at that time the server is empty, then service will be started for class-1 flit, as it has higher priority.
Therefore, the portion of residual time that occurs due to class-1 flits will decay linearly from $T_i$ instead of $(T_i - 1)$.
Therefore, $R^{(2)}$ can be written as:
\begin{align} \nonumber
R^{(2)} &= \frac{1}{T_{tot}}\sum_{i=1}^{M_1(T_{tot})} \Bigg( \sum_{\tau=0}^{T_i} \tau \Bigg) + \frac{1}{T_{tot}} \sum_{j=1}^{M_2(T_{tot})}\Bigg( \sum_{\tau=0}^{T_j-1} \tau\Bigg) \\ \nonumber 
&= \frac {1}{2} \lambda_1 (\overline{T_1^2}+\overline{T_1}) + \frac {1}{2} \lambda_2 (\overline{T_2^2}-\overline{T_2}) \\ \nonumber 
&= \frac {1}{2} \lambda_1 (\overline{T_1^2}-\overline{T_1}) + \lambda_1 \overline{T_1} + \frac {1}{2} \lambda_2 (\overline{T_2^2}-\overline{T_2}) \\  \label{eq_R_2}
&= R_1 + \rho_1 + R_2
\end{align}
In general, if there are total N classes, residual time expression for flits of class-$i$ will be:
\begin{equation}
R^{(i)} = 
\begin{cases}
\sum_{k=1}^{N}R_k, & \text{for $i = 1$ } \\
\\
\sum_{k=1}^{N}R_k + \sum_{k=1}^{i-1} \rho_k,  & \text{for $i > 1$}
\end{cases}
\end{equation}
where $R_k = \frac{1}{2} \lambda_k (\overline{T_k^2}-\overline{T_k})$ for Geo/G/1 queues and $\rho_k$ is the average utilization of server for class-$k$ flits.

\vspace{1mm}

\noindent\textbf{Average Queuing Time Expressions: } \label{sec: waiting time expressions}
The queuing time expression for tokens with traffic class-1 and class-2 can be written as~\cite{bertsekas1992data}:
\begin{equation}
W_1 = R^{(1)} + \rho_1 W_1
\implies W_1 = \frac{R^{(1)}} {1-\rho_1}
\end{equation}
\begin{equation}
W_2 = R^{(2)} + \rho_1 W_1 + \rho_2 W_2 + \rho_1 W_2 \\
\implies W_2 = \frac{R^{(2)} + \rho_1 W_1}{1-\rho_1-\rho_2}
\end{equation}

Substituting $R^{(1)}$ and $R^{(2)}$ from Equation \ref{eq_R_1_final} and \ref{eq_R_2} respectively:
\begin{align}
W_1 &= \frac{R_1 + R_2} {1-\rho_1} \\
W_2 &= \frac{R_1 + \rho_1 + R_2+ \rho_1 W_1}{1-\rho_1-\rho_2}
\end{align}

In general, if there are total N classes, queuing time expression for class-$i$ flits will be
\begin{equation}
W_i = 
\begin{cases}
\frac{\sum_{k=1}^{N}R_k} {1-\rho_1}, & \text{for $i = 1$ } \\
\frac{\sum_{k=1}^{N}R_k + \sum_{k=1}^{i-1} \Big(\rho_k + \rho_k W_k \Big)}{1 - \sum_{k=1}^{i} \rho_k}, & \text{for $i > 1$}
\end{cases}
\end{equation}


\end{document}